\documentclass[12pt,3p,twocolumn]{elsarticle}

\usepackage{amssymb}
\usepackage[switch]{lineno}

\usepackage{graphicx}
\usepackage{amsmath}
\usepackage{mathtools}
\usepackage{times}
\usepackage{bm}
\usepackage{natbib}
\usepackage{url}
\usepackage{multirow}
\usepackage{tabularx, ragged2e, booktabs}
\newcolumntype{Y}{>{\centering\arraybackslash}X}

\usepackage{array}
\newcolumntype{L}[1]{>{\raggedright\let\newline\\\arraybackslash\hspace{0pt}}m{#1}}
\newcolumntype{C}[1]{>{\centering\let\newline\\\arraybackslash\hspace{0pt}}m{#1}}
\newcolumntype{R}[1]{>{\raggedleft\let\newline\\\arraybackslash\hspace{0pt}}m{#1}}

\usepackage[page]{appendix}

\newcommand{\hl}{} % instead of {\bf } write only {} 
\newcommand{\hll}{} % instead of {\bf } write only {} 

\journal{Physica D}

\begin{document}

%\linenumbers

\begin{frontmatter}



\title{Predictability of threshold exceedances in dynamical systems}


\author{Tam\'as B\'odai\corref{cor1}}
\address{Meteorological Institute, University of Hamburg, Grindelberg 5, 20144 Hamburg, Germany} 
\cortext[cor1]{tamas.bodai@uni-hamburg.de, +49 40 42838 9205}
%\ead{tamas.bodai@uni-hamburg.de}

\begin{abstract}

In a low-order model of the general circulation of the atmosphere we examine the predictability of threshold exceedance events of certain observables. The likelihood of such binary events {\hl-- the cornerstone also for the {\em categoric} (as opposed to probabilistic) prediction of threshold exceedences --} is established from long time series of one or more observables of the same system. The prediction skill is measured by a summary index of the ROC curve that relates the hit- and false alarm rates. Our results for the examined systems {\hll suggest that {\em exceedances of higher thresholds are more predictable}; or in other words: rare large magnitude, i.e., extreme, events are more predictable than frequent typical events. We find this to hold provided that the bin size for binning time series data is optimized, but not necessarily otherwise. This can be viewed as a confirmation of a counterintuitive (and seemingly contrafactual) statement that was previously formulated for more simple autoregressive stochastic processes. However, we argue that for dynamical systems in general it may be typical only, but not universally true.} We argue that when there is a sufficient amount of data depending on the precision of observation, the skill of {\hll a class of} data-driven {\hl categoric predictions of threshold exceedences} approximates the skill of {\hl the {\hll analogous} model-driven} prediction, assuming strictly no model errors. {\hl T}herefore, stronger extremes {\hl in terms of higher threshold levels} are more predictable {\hl both in case of data- and model-driven prediction}. Furthermore, we show that a quantity commonly regarded as a measure of predictability, the finite-time maximal Lyapunov exponent, does not correspond directly to the ROC-based {\hl measure of prediction skill} when they are viewed as functions of the prediction lead time and the threshold level. This points to the fact that even if the Lyapunov exponent as an intrinsic property of the system, measuring the instability of trajectories, determines predictability, it does that in a nontrivial manner.

\end{abstract}


\end{frontmatter}

\section{Introduction}

Extreme events have fundamental importance to life, as they are often associated with survival and losses. -- Extreme events to do with gains or amusement receive far less attention in general, dissociated from individual events. Rare and large magnitude events of interest arise in physical, technological, social, and other systems~\cite{npg-18-295-2011}. The classical theory of extremes in uncorrelated sequences (or, in sequences in which the auto-correlation is decaying sufficiently fast)~\cite{FT:1928,Gnedenko:1943,LLR:1983} has a statistical orientation; it is not- and cannot be concerned with {\em prediction} or with uncovering mechanisms that can produce extremes; but it is rather concerned with e.g. expected return times, which can be useful in designing structures of a certain required life time, such as sea walls~\cite{Coles:2001}. 

Since Newton revolutionized science, it has become a paradigm that predictions should be based on validated models. These models describing fluctuating phenomena often take the form of a system of differential equations, also referred to as a {\em dynamical system}. Since the work of Lorenz it has become clear that even though some phenomena can be modeled quite accurately, they can be inherently unpredictable because of their extreme sensitivity to initial conditions~\cite{Tel_n_Gruiz:2006}. Such systems are called chaotic, characterized by positive Lyapunov exponents. This imposes a time horizon on predictions; beyond that only statistical properties can be robustly estimated{\hl, which is what classical extreme value theory is concerned with. In contrast with that, in our analysis we consider prediction lead times shorter than the decorrelation time in a time series.}  

{\hll In the context of} model-based or {\em model-driven} predictions (MDP) {\hll equivalent with initial value problems for deterministic differential equations}, like e.g. a weather forecast, one can often read that extremes are much harder to predict. Unfortunately a systematic study of the dependence of some appropriate {\hl prediction} skill score {\hl -- or a measure of predictability in a more general sense --} of any model on the magnitude of events is still lacking. Inaccuracy of the model may be an important factor {\hll leading to} such a dependence {\hl of its prediction skill on the event magnitude}, beside details of its chaotic nature. In contrast, in pure {\em data-driven} prediction (DDP), model errors are not present, {\hl as the basis of the prediction (of any kind) does not involve a model in the form of equations or an algorithm, only observational/measurement data}. Instead, beyond errors in measuring the present conditions (as with initial conditions for MDP), the prediction is compromised by the finite size of the data set. That is, the said virtue of DDP can be exploited -- when employing it in its pure form -- only if enough and good quality data (with a high precision of observation and high signal-to-noise ratio) is available~\cite{Bogachev20112240}.
 
One might expect that the slogan that `extremes in comparison with more moderate events are harder to predict' extends to DDP. In fact, just the opposite has been reported by Hallerberg and Kantz~\cite{npg-15-321-2008} for simple autoregressive processes at least, indifferently to whether the probability distribution is exponentially decaying or according to a power-law, and also for some observational data~\cite{10.1371/journal.pone.0111506}: {\em stronger events are easier to predict}. This counter-intuitive statement is based on a {\hl measure of prediction skill} that derives from the so-called {\em receiver operating characteristic} (ROC) {\em curve}~\cite{HBK:2008,Egan:1975} {\hll that} takes into account the true positives -- meaning that an event is correctly predicted to happen -- as well as the false negatives. Concerning rare events, such a {\hl measure of prediction skill} is regarded~\cite{Kantz:2010} more meaningful than other {\em proper}~\cite{BS:2007} so-called {\em skill scores} for probabilistic predictions like the Brier or Ignorance scores. {\hl This is so, because t}he ROC statistics {\hl has been viewed to} not depend on the relative frequency of events (only that the accurate evaluation of the statistics requires a sufficient number of events). {\hl {\hll The latter} characteristic is thought to allow for the comparison of the ROC-predictability of events between two situations where the events have different frequency~\cite{Kantz:2010}}. 

Whether the above statement~\cite{npg-15-321-2008} can be {\em maintained} in case of more complex processes has been an open question so far -- addressed but not settled with a consensus. Recently two studies~\cite{PhysRevE.85.031134,npg-19-529-2012} have been published concerning the predictability of extreme events in dynamical systems with seemingly contradictory results as to whether stronger events are more predictable. Franzke~\cite{PhysRevE.85.031134} {\hl applied the method set out in~\cite{HBK:2008} to predict extreme threshold exceedances} in a systematically derived stochastic dynamical system representing climate variability by the resolved (slow) variable(s) and weather variability by noise in place of the unresolved (fast) variable(s)~\cite{Majda10032009}. He maintained the earlier statement~\cite{npg-15-321-2008} in this case, measuring the prediction skill by the ROC statistics, but on the basis of {\hll considering only two high threshold values}. On the other hand, Sterk et al.~\cite{npg-19-529-2012} considered a number of dynamical systems of various complexity, and various physical observables. They evaluated finite-time maximal Lyapunov exponents {\hll (FTMLE)} {\hl of} trajectories that lead to extremes, and concluded that no generally applicable statements can be made, but the predictability of extremes depends on the system (and so the attractor geometry) and on the observable in question, as well as the prediction lead time. We emphasize that in their study the authors did not take model errors into account. 

{\hl To summarize the essence of the above review, we can list three different views encountered in the literature regarding the predictability of extremes:
\begin{description}
 \item[(1)] Stronger extremes are better predictable.
 \item[(2)] Stronger extremes are less   predictable.
 \item[(3)] Stronger extremes can be better or less predictable depending on various factors.
\end{description}
Without giving details, {\hll e.g. assumptions of these statements, they} seem to be contradictory to each other. On this basis we set out the following objectives for the present paper:
{\hll 
\begin{description}
 \item[(i)] Keeping to the assumption of (1), we evaluate the predictability of {\em peak-over-threshold} events measured by a ROC-based quantity, using time series data of finite length produced by the Lorenz-84 model~\cite{L84}. With an attention to (3), we evaluate (i.a) the dependence of predictability {\em itself} on various factors, and also what is more relevant to the question: (i.b) the magnitude-dependence of predictability -- whether increasing/decreasing or nonmonotonic -- depending on some of those same factors.
 \item[(ii)] We argue for an analogy between a certain class of DDPs and MDP, and that the latter is usually understood as something that below we will refer to as an {\em on-demand} MDP, in which case any input data belongs to a single time instant. We believe this is an assumption of (2). This objective (ii) is to reconcile (1) and (2), suggesting that (2) can be true when model errors are present, even if the predictability is measured by the same ROC-based quantity as that assumed by (1).
 \item[(iii)] For on-demand MDP and the analogous DDP where the time of input is arbitrary, we will be able to carry out an assessment of the lead time-dependence of the predictability of what we will call {\em threshold-exceedance-in-an-interval} events in a straightforward manner. This will turn out to have a bearing on the magnitude-dependence of predictability. This objective (iii) together with (i.b) are to revisit point (1), possibly extending that point from stochastic processes to dynamical systems.
 \item[(iv)] However, to show that (3) does not necessarily contradict (1), we recall that (3) was stated on the basis of measuring predictability by FTMLEs. This also assumes an arbitrary input time on-demand MDP or analogous DDP. Accordingly, in the autonomous L84 we calculate the FTMLEs of trajectories that lead to extremes, and compare their average to the ROC-based measure of prediction skill -- looking for any qualitative mismatch. 
\end{description}
}

To motivate our top objective (i) {\hll and (iii)} we remark that} DDP is gaining increasing prominence nowadays given that data is relatively much more easily accessible than models. This is certainly the case with geophysical phenomena that we are primarily interested in, such as meteorology. Furthermore, performing predictions based on data can be far less costly than those based on simulating complex models, while the skill may not be much worse~\cite{SBK:2013}.

Next we recapitulate the methodology of the applied prediction scheme and the used ROC-based {\hl measure of prediction skill}. Lorenz's 1984 model of global atmospheric circulation, simulated to produce time series data for the purpose of assessing the predictability of threshold exeedance events, is also briefly described. Subsequently, in Sec. \ref{sec:results}, we present our results on the dependence of predictability on several factors, such as: {\hl the makeup of the so-called precursory structure -- made use for a prediction -- in terms of the observables involved}, the prediction lead time, and the magnitude of extreme events. These results {\hll pertaining to objectives (i) and (iii)} are summarized in a compact table format in Sec. \ref{sec:summary} and discussed subsequently {\hl regarding {\hll objectives (ii) and (iv)}}. To close our presentation we pose a few open questions for future research into practical aspects of the prediction of extremes, which might potentially have theoretical ramifications. {\hl We provide in Appendices \ref{sec:max_find_alg} and \ref{sec:pred_model_driv}, respectively, the description of an algorithm for finding the maximum of a function of one variable and a definition of the finite-time Lyapunov exponents.}

\section{Methodology}\label{sec:methodology}

\subsection{Prediction of threshold exceedances by precursors}\label{sec:pred_data_driv}

\begin{figure*} %[t!]
    \begin{center}
	  \includegraphics[width=\linewidth]{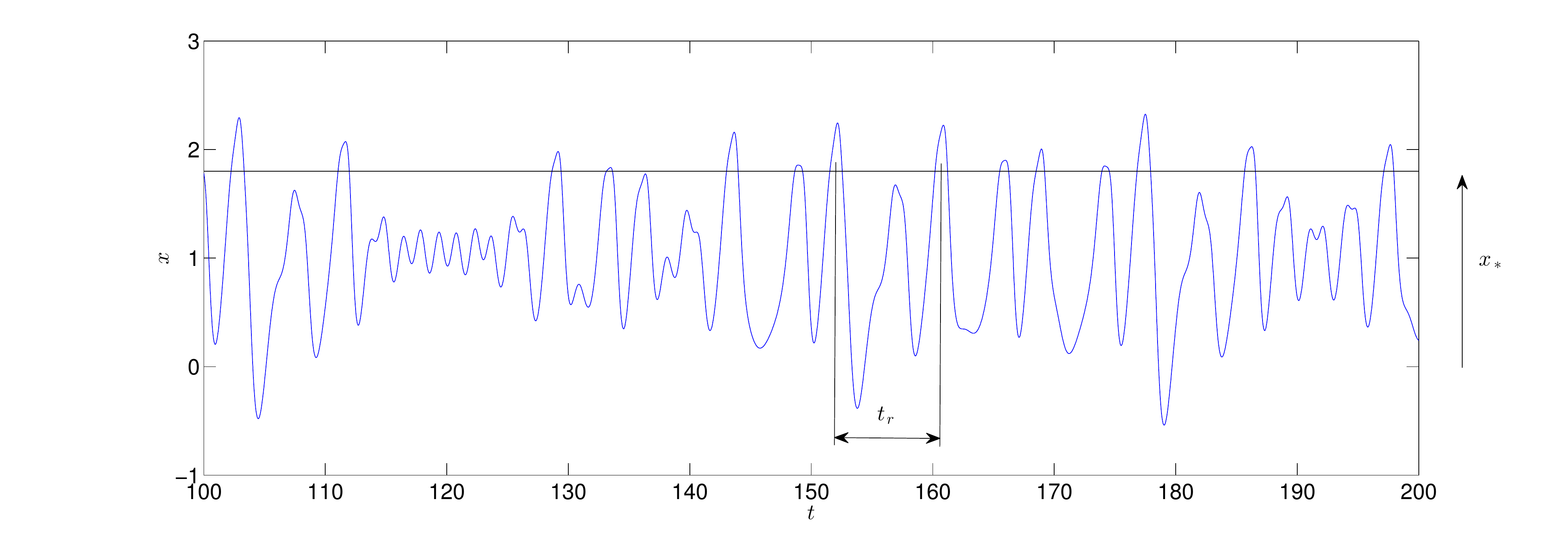} 
        \caption{\label{fig:time_history} Illustration of the prediction problem: given a continuous-time chaotic time series $x(t)$ featuring apexes that every so often overshoot a relatively high threshold $x_*$ (straight horizontal line), we want to predict these overshoots, i.e., extreme events, at the time, say, of the immediately preceding apex. The time series shown was in fact generated by simulating the Lorenz 84 model (\ref{eq:L84}) with $F=8$, and we plotted the first component of the solution $x$. The time between two subsequent apexes above the threshold is called the return time $t_r$ of extremes.
        }
    \end{center}
\end{figure*}

Our aim is to predict large excursions of some (scalar) physical observable $x$, exceeding a chosen threshold level $x_*$, before that exceedance happens. Figure \ref{fig:time_history} pictures the situation as the observable evolves {\em continuously} in time, $x=x(t)$, $t\in\mathbb{R}$, occasionally exceeding the threshold. We intend to examine situations when $x(t)$ is generated by a process that can be described by an ordinary or stochastic differential equation, examples of which for our case study will be briefly described in Sec. \ref{sec:model}. {\hl When one does not have a validated model, only the time series $x(t)$, one can still perform DDP; when the opposite is the case, one can do MDP.}

\subsubsection{Formal setting}\label{sec:formal_setting}

%{\em Formal setting.}
The following methodological description regarding the prediction task closely follows~\cite{HBK:2008}. {\hl It applies to DDP as well as MDP, on which we comment in the end of this section.} We introduce a discrete-time {\em binary} event variable:

\begin{equation}\label{eq:chi}
   \chi_n = \left\{ 
   \begin{tabular}{cc}
      1, & $x(t_n) > x_*$ \\
      0, & $x(t_n) < x_*$
   \end{tabular}
   \right.
\end{equation} 
where the times $t_n$, $n\in\mathbb{Z}$, belong to consecutive apexes, that is, local maxima or peaks, of the continuous evolution of $x$ for consecutive values of $n$, and in general they are not equally `spaced'. Thereby the continuous-time evolution is {\hl{\em discretized}. (This can be viewed as a restriction on generality, and it will be relaxed further below.)} This approach is often referred to as the peak-over-threshold (POT) approach. The prediction is based on a like-wise discrete-time precursory structure $\mathbf{x}_n\in\mathbb{R}^M$ of size $M$, whose different members, observables desirably related to $x$, may belong to different times, e.g. $t_{n-d_m}$, preceding the current time $t_n$, specified by delays $d_m\in\mathbb{Z}$, $m=1,\dots,M$. We call $t_n-t_{n-\min(d_m)}>0$ the prediction lead time. Our {binary} prediction for $\chi_n$ at $t_{n-\min(d_m)}$ is defined as:  

\begin{equation}\label{eq:chi_hat}
   \hat{\chi}_n = \left\{ 
   \begin{tabular}{cc}
      1, & $\mathcal{L}(\mathbf{x}_n) > \mathcal{L}_*$ \\
      0, & $\mathcal{L}(\mathbf{x}_n) < \mathcal{L}_*$
   \end{tabular}
   \right.
\end{equation} 
based on the {\em likelihood} function:

\begin{equation}\label{eq:likelihood}
   \mathcal{L}(\mathbf{x}) = \mathbb{P}_{\chi|\mathbf{x}}(\chi = 1,\mathbf{x}) = \mathcal{P}(\mathbf{x}) / p(\mathbf{x}).
\end{equation} 
In the above $\mathcal{P}(\mathbf{x})=p_{\mathbf{x}| \chi}(\mathbf{x} , \chi = 1)\mathbb{P}_{\chi}(\chi = 1)$ is the {\em posterior} probability density function (PDF) of $\mathbf{x}$, and $p(\mathbf{x})$ is the {\hl `process'} PDF, i.e., the basic PDF generated by the considered process\footnote{\hll The probability density $p_{\mathbf{x}| \chi}(\mathbf{x} , \chi = 1)$ of $\mathbf{x}$ conditioned on some realized value of $\chi$ is usually denoted more simply as $p(\mathbf{x} | \chi = 1)$, but we want to emphasize that we consider a function of two variables. Also, it would create ambiguity if the symbol $p$ without a subscript was to be reused to denote another function, and we prefer to reserve $p$ for the process PDF.}. {\hl Refer to the appendix of~\cite{PhysRevE.77.011108} for an integral formulation of e.g. $\mathcal{P}(\mathbf{x})$ which applies the Heaviside step function as a filter.} Note that Eq. (\ref{eq:likelihood}) expresses Bayes' theorem relating the conditional probabilities: the likelihood and the posterior probability. Our prediction $\hat{\chi}_n$ is controlled by a threshold $\mathcal{L}_*\in[\min(\mathcal{L}),\max(\mathcal{L})]$ of stringency on $\mathcal{L}$. Note that an actual choice is meant to be made as to the applied value of $\mathcal{L}_*$ in practice, for which reason this kind of prediction is not probabilistic, {\hl but we call it a categoric prediction}. 

\begin{figure} %[t!]
    \begin{center}
	  \includegraphics[width=\linewidth]{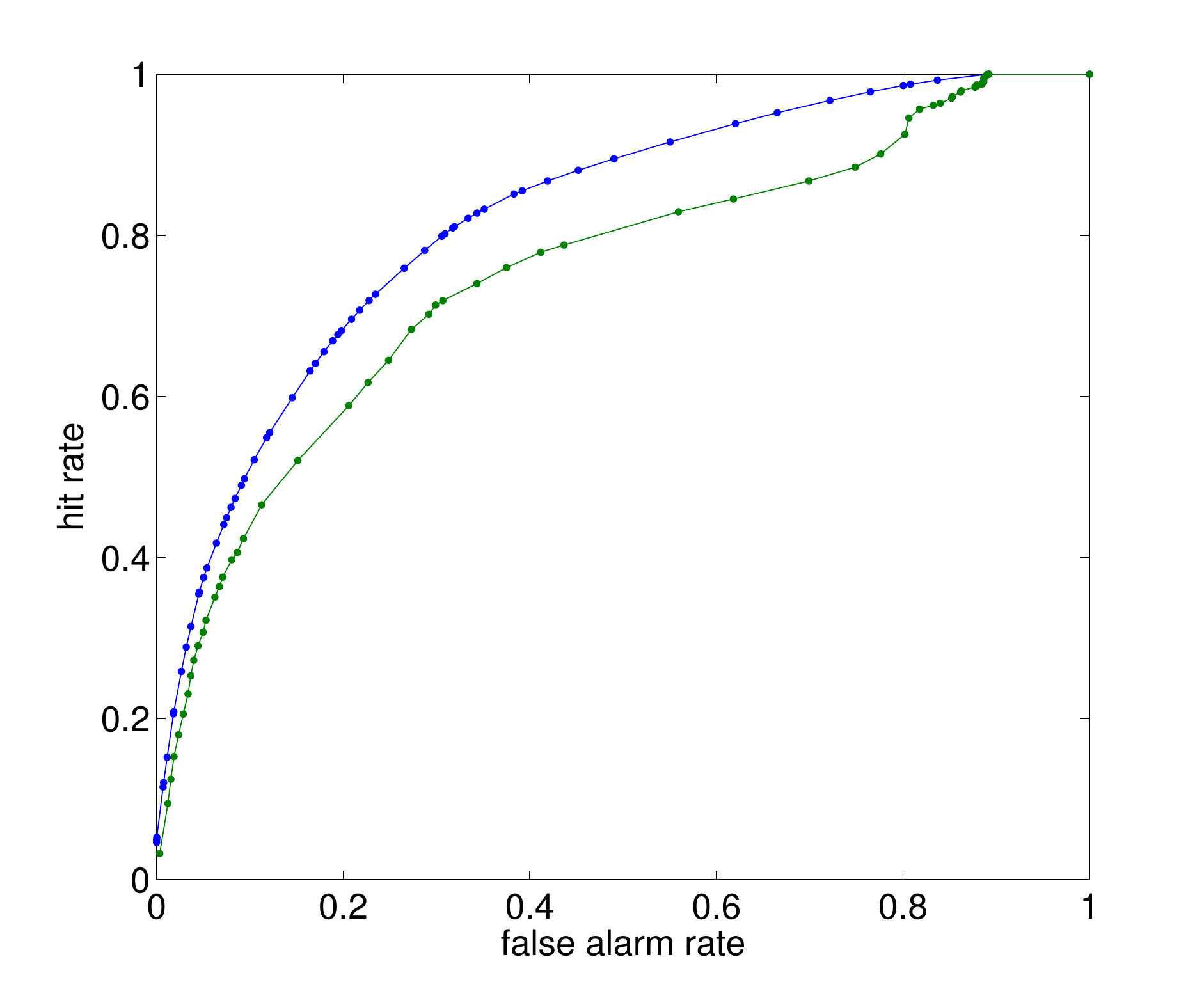} 
        \caption{\label{fig:ROC_Pp_L_x} Example $\mathcal{L}$- (above, blue) and $\mathcal{P}$-ROC curves (below, green) corresponding to the scenario of the autonomous L84 in Fig. \ref{fig:time_history} when $\mathbf{x}_n=x_{n-1}$ {\hl and $x_*=1.8$. The distributions that these ROC curves derive from can be seen in Fig. \ref{fig:Pp_L_x} drawn by the thin black line.}}
    \end{center}
\end{figure}

Depending on $\mathcal{L}_*$, the rate of true positives, or the {\em hit rate}, i.e., the frequency of making correct predictions, yields as follows: 

\begin{equation}\label{eq:hit_rate}
  H(\mathcal{L}_*) = \frac{ \int_{\mathbb{R}^M}dV_{\mathbf{x}} \mathcal{P}(\mathbf{x})\mathcal{H}(\mathcal{L}(\mathbf{x})-\mathcal{L}_*) }{ \int_{\mathbb{R}^M}dV_{\mathbf{x}} \mathcal{P}(\mathbf{x}) }.
\end{equation} 
{\hl In the above $dV_{\mathbf{x}}$ is a volume element in the precursory space, and $\mathcal{H}(\cdot)$ is the Heaviside step function.} Another measure of the overall goodness or skill of prediction is the {\em false alarm rate}: 

\begin{equation}\label{eq:far}
  F(\mathcal{L}_*) = \frac{ \int_{\mathbb{R}^M}dV_{\mathbf{x}} [p(\mathbf{x}) - \mathcal{P}(\mathbf{x})]\mathcal{H}(\mathcal{L}(\mathbf{x})-\mathcal{L}_*) }{ \int_{\mathbb{R}^M}dV_{\mathbf{x}} [p(\mathbf{x}) - \mathcal{P}(\mathbf{x})] }.
\end{equation}
Clearly, one can achieve a very good hit rate by reducing the stringency, but in fact~\cite{HBK:2008} always at the price of an increased false alarm rate. Figure \ref{fig:ROC_Pp_L_x} shows an example of how the two measures of skill depend on the stringency in terms of a parametric plot or curve $\{(F(\mathcal{L}_*),H(\mathcal{L}_*))\}$, which is referred to as the {\em receiver operating characteristic} (ROC) {\em curve}. With the extremal choices, $\mathcal{L}_*=0$ and 1, we have $(F=1,H=1)$ and $(F=0,H=0)$, respectively, i.e., the ROC curve stretches from corner to corner. It is a diagonal straight line with no prediction skill at all {\hl(over random predictions $\hat{\chi}$ with $\mathbb{P}(\hat{\chi} = 1)=\mathcal{L}_*$)}, and situated above the diagonal with any skill. In the same diagram another ROC curve is also shown, to be referred to as $\mathcal{P}$-ROC curve, obtained by {\hl writing within the scope of the Heaviside function in the definitions (\ref{eq:hit_rate}), (\ref{eq:far}), and also in (\ref{eq:chi_hat}), $\mathcal{P}$ instead of $\mathcal{L}$ and $\mathcal{P}_*$ instead of $\mathcal{L}_*$. Note that $\mathcal{P}_*\in[\min(\mathcal{P}),\max(\mathcal{P})]$.} This is based on the intuitive strategy, expressed by the conditional probability $p_{\mathbf{x}| \chi}(\mathbf{x} , \chi = 1)$, that one looks at what happens before extreme events. From Eq. (\ref{eq:likelihood}) one can see that following this strategy the posterior PDF is just the likelihood that such states lead to an extreme event weighed by the relative frequency of those states, whereby the `predictive potential of relatively infrequent states is suppressed'. It can be shown~\cite{HBK:2008,10.1371/journal.pone.0111506} that as a result of this the $\mathcal{P}$-ROC curve will be always wholly underneath the $\mathcal{L}$-ROC curve, making this intuitive strategy inferior. Furthermore, the $\mathcal{L}$-ROC curve is always concave~\cite{Egan:1975}, while the $\mathcal{P}$-ROC curve is not necessarily concave. Besides, in accordance with the above statement on the trade-off situation, $F_{\mathcal{L}}(\mathcal{L}_*)$, $H_{\mathcal{L}}(\mathcal{L}_*)$, $F_{\mathcal{P}}(\mathcal{P}_*)$, $H_{\mathcal{P}}(\mathcal{P}_*)$ are all monotonic functions, and, therefore, so are e.g. $H_{\mathcal{L}}(F_{\mathcal{L}})$ and $H_{\mathcal{P}}(F_{\mathcal{P}})$.

The ideal situation when extreme events ($\chi=1$) and nonevents ($\chi=0$) can be predicted with certainty ($\hat{\chi}=\chi$) is represented by the $(F=0,H=1)$ corner in the ROC diagram. In this case no choice has to be made on the applied stringency $\mathcal{L}_*$. Clearly this is possible only in case of the deterministic but not the stochastic version of a model, and there are further factors -- to be demonstrated in Sec. \ref{sec:results} -- that can compromise the prediction skill. In the nonideal situation an optimal $\mathcal{L}_*$ is to be chosen. A unique optimum exists only in terms of a single-objective optimization problem, defined by a scalar-valued cost function. However, in our case the minimization of the false alarm rate and the maximization of the hit rate are both `valid' objectives. It takes a {\em specific application} to be possibly able to define a scalar-valued cost function $C(F,H)$. For our {\em general assessment} of predictability we choose to consider the intuitive measure:

\begin{equation}\label{eq:distance}
   D = \min_{\mathcal{L}}(\sqrt{F^2+(H-1)^2}),
\end{equation} 
the distance of the ROC curve from the ideal corner. With no prediction skill at all: $D=\sqrt{2}/2$. {\hl Other summary statistics for the ROC curve have also been defined, such as the area under the curve~\cite{HBK:2008}, or the slope $H'_F(F=0)$~\cite{npg-15-321-2008}. Unlike these two, the distance $D$ can be associated to actual predictions specified by an actual choice for $\mathcal{L}_*$.}
We note that it is not trivial to interpret what the comparison of $D$ with a proper skill score of probabilistic prediction means.

\subsubsection{Numerical issues}\label{sec:numerics}

%{\em Numerical issues.}
Perhaps the most obvious factor that compromises the prediction skill in the data-driven framework is the finite size $N$ of the data set: $\{x_n,\mathbf{x}_n\}$, $n=1,\dots,N$. The distributions $p(\mathbf{x})$, $\mathcal{P}(\mathbf{x})$, $\mathcal{L}(\mathbf{x})$ will be approximated in our study by histograms $\{{p}_b\}$, $\{{\mathcal{P}}_b\}$, $\{{\mathcal{L}}_b\}$, $b=1,\dots,B$, of a certain uniform bin size $\Delta\mathbf{x}\in\mathbb{R}^M$; different values of $b$ can be assigned to the different bins by a sensible algorithm. Let us denote by $\Delta x$ the unique linear bin size applied in all $M$ dimensions after suitable nondimensionalization. Note that in the coarse-grained situation $\{{\mathcal{L}}_b\}$ derives from $\{{p}_b\}$ and $\{{\mathcal{P}}_b\}$ much the same way as with the continuous functions according to Eq. (\ref{eq:likelihood}). With the discrete formulation of Eqs. (\ref{eq:hit_rate}) and (\ref{eq:far}), accordingly, the ROC curve turns into (the graph of) a staircase (function), given by a set of discrete data points: $\{({H}_b,{F}_b)\}$, $b=1,\dots,B$, belonging to stringency levels $\{\mathcal{L}_{*,b}\}=\{\mathcal{L}_b\}$: 

\begin{eqnarray}
  H_b &=& \frac{\sum_{b'=1}^B\mathcal{P}_{b'}\mathcal{H}(\mathcal{L}_{b'}-\mathcal{L}_b)}{\sum_{b'=1}^B\mathcal{P}_{b'}} \label{eq:hit_rate_discr} \\ 
  F_b &=& \frac{\sum_{b'=1}^B(p_{b'}-\mathcal{P}_{b'})\mathcal{H}(\mathcal{L}_{b'}-\mathcal{L}_b)}{N-\sum_{b'=1}^B\mathcal{P}_{b'}} \label{eq:far_discr}
\end{eqnarray}
{\hl Note that in the above the histograms $\{{p}_b\}$ and $\{{\mathcal{P}}_b\}$ do not need to be normalized; and e.g. $\{{\mathcal{L}}_b\}$ and $\{{\mathcal{L}}_{b'}\}$, $b,b'=1,\dots,B$, denote the same set.} Note {\hl also} that if $\mathcal{L}_b$ does not-, then neither do ${H}_b$ and ${F}_b$ change monotonically with increasing $b$. 

The above estimation of the measures of skill is not conservative\footnote{\hll By `conservative estimation of the skill' we mean that an overestimation of skill (e.g. $D$ estimated to be smaller than the true value) is excessively unlikely. This entails an appropriate sign of the bias of the estimator, and a standard deviation of the estimator much smaller than the modulus of the bias.}, however, which is to do with small histogram counts and associated statistical errors. An approach to fix this problem is the following. The available data is divided equally into `training' and `evaluation' data sets. Then, the conservative estimates are defined again by Eqs. (\ref{eq:hit_rate_discr}) and (\ref{eq:far_discr}), but the different terms appearing in them are associated with different data sets: $\{{\mathcal{L}}_b\}$ is derived from the training data set, and $\{{p}_b\}$ and $\{{\mathcal{P}}_b\}$ are derived from the evaluation data set. {\hl Note that the latter requires the use of the same grid forming the bins in case of the training and evaluation data sets.}
 
A further issue to do with small bin sizes when many bins contain a single data point is that the `ROC staircase' can have an excessively large {\em last} step. This is so, because bins that contain single data points tend to have empty counterparts mutually between the `training' and `evaluation' data sets. This way $D>\sqrt{2}/2$ can even be realized.
 
Too large bin sizes would of course also deteriorate the prediction skill. Therefore, there should be an {\em optimal} bin size yielding (locally) minimal $D$. Our numerical experience shows that there is always, for any given prediction lead time or threshold level $x_*$, a unique (globally) optimal {\em uniform} bin size defining the {\em regular} grid.

\subsubsection{Relationship of data- and model-driven predictability}

%{\em Relationship of data- and model-driven predictability.} 
We note finally that the above description of evaluating predictability applies clearly to the case of data-driven prediction. However, evaluating the model-driven predictability of binary exceedance events measured by the same ROC-based {\hl measure of prediction skill} can be done in exactly the same way: by simulating long trajectories of the model and by binning the resulting time series data. This is {\hl more obviously true} when MDP is thought of in the sense that the {\hll complete} PDFs{\hl, $p(\mathbf{x})$, $\mathcal{P}(\mathbf{x})$, $\mathcal{L}(\mathbf{x})$, or rather the histograms, $\{{p}_b\}$, $\{{\mathcal{P}}_b\}$, $\{{\mathcal{L}}_b\}$,} are established preliminary to making any predictions. {\hll We will refer to such an MDP as an `archival' MDP. Note that in this sense any DDP is archival. The} archival MDP and {\hll the corresponding} DDP differ only in that in principle unlimited data is available to establish the PDFs for MDP. But MDP can be thought of also in the sense that the model is only simulated `on-demand', to produce an ensemble forecast from which the likelihood of a threshold exceedance can be established. {\hl The case of on-demand MDP corresponds to {\hll imposing a constraint on} archival MDP or DDP}, {\hll namely that} the components of the precursory structure $\mathbf{x}$ of the archival MDP or DDP belong to the same time instant ($d_m=d_{m'}$, $m,m'\in[1,...,M]$), {\hl$\mathbf{x}$ thereby} representing the {\em initial conditions} for the on-demand MDP. Note that it is allowed that $\mathbf{x}$ excludes some variables that determine the considered phenomenon {\hl($M<d$, $d$ being the dimension of the phase space, as detailed in Appendix \ref{sec:pred_model_driv}); the excluded variables} can be initialized arbitrarily, or possibly as a random sample from a probability distribution. We refer here to the practice of stochastic parametrization of unresolved processes in weather forecast models. 

As for the definition of the binary event in case of on-demand MDP the following can {\hll also} be taken: in an event a chosen observable exceeds a set threshold in a chosen future time interval, defined by a leading window of width $\Delta T$ at a lead time $T$ ahead of the present time. Let us call this an event of {\em threshold exceedance in an interval} (TEI event in short). In contrast to considering the threshold exceedance of apexes of the time evolution, i.e., POT events, here $T$ can be set arbitrarily, not restricted to discrete values; accordingly, $\chi(t)$ and $\mathbf{x}(t)$ are defined in continuous-time. {\hl We will take this approach when assessing predictability with respect to the prediction lead time in Sec. \ref{sec:lead_time}.}

The bin size $\Delta x$ {\hll for a DDP} would in this case {\hl correspond roughly to} the precision $\delta x$ of measuring initial conditions {\hll for a MDP}. {\hl(The difference is that $\Delta x$ corresponds to averaging of the distributions, e.g. $\mathcal{L}(\mathbf{x})$, over predefined fixed bins, whereas $\delta x$ corresponds to a smoothing of the distribution: $\tilde{\mathcal{L}}(\tilde{\mathbf{x}})=\int d\mathbf{x}\mathcal{L}(\mathbf{x})p_e(\tilde{\mathbf{x}}-\mathbf{x})$, where $\delta x$ can be thought of as the standard deviation of the measurement error distribution $p_e(\tilde{\mathbf{x}}-\mathbf{x})$.)} In order to well-approximate the skill of MDP for a given measurement precision, the time series has to be long enough so that the likelihood for all bins, even those which cover a relatively small portion of the invariant measure of the attractor, is well-approximated on the first place. That is, the ratio of the prediction skill of DDP and that of MDP (with no model errors), which is always smaller than unity, depends on $N\Delta x$.

\subsection{The model climate}\label{sec:model}

We carry out the assessment of the predictability of extremes in a model of geophysical relevance. {\hll It} constitutes a nonlinear dynamical system featuring complex chaotic deterministic dynamics. {\hll Yet, it} is simple enough -- involving just three scalar prognostic variables -- to yield {\hl time series} data relatively inexpensively, and to allow for a more tangible demonstration of some aspects of predictability. Our choice of a model, Lorenz's model of global atmospheric circulation (L84) with standard parameter settings, reads as follows~\cite{L84}:

\begin{equation}\label{eq:L84}
    \begin{aligned}
        \dot{x}&=-y^2-z^2\ -x/4\ + F/4,\\
        \dot{y}&=xy-4xz\ -y +1,\\
        \dot{z}&=xz+4xy\ -z.
    \end{aligned}
\end{equation}
The model describes -- in a very coarse manner~\cite{Veen:2003} -- the meridional heat transport via eddies, represented by principal mode amplitudes $y$ and $z$, given rise by the baroclininc instability of the midlatitude jet, represented by its average speed $x$. The instability occurs for appropriate conditions defined by the large scale meridional temperature gradient, represented in the model by $F$, due to differential heating between the equator and the poles. The equations are nondimensionalized with respect to time by the average damping time of eddies, being about 5 days. {\hl This model enjoys popularity in teaching ~\cite{Provenzale:1999,Tel_n_Gruiz:2006} as well as theoretically oriented weather and climate research~\cite{Masoller1995357,QJ:QJ49712152312,PhysRevE.87.022822,DBT:2015}.} 

We will examine the autonomous dynamics in perpetual winter conditions realized by, say, $F=8$, since this gives rise to chaotic dynamics, which is nontrivial from the point of view of predictability. {\hl Let us label the autonomous dynamics/model by M1, as a synonym of (\ref{eq:L84}), when the forcing takes the above indicated form.} {\hll In order to assess the dependence of predictability on intrinsic system properties, according to objective (i.a),} we will consider also nonautonomous dynamics. {\hll It is achieved} by introducing some driving or time-dependent forcing to the L84 system in the form of $F(t)=F_0+A\tilde{x}(t)$, where the fluctuating process (in a mathematical sense) $\tilde{x}(t)$ can be seen as an unresolved, i.e., physically not modeled, (physical) process. However, from the point of view of data-driven predictability, as described in Sec. \ref{sec:pred_data_driv}, autonomous and nonautonomous systems are not distinguished -- the {\em driving mimics additional degrees of freedom} of the system. We can represent additional degrees of freedom of {\em comparable} time scales to that of the resolved dynamics, $\tau_{L84}\approx4$, by a continuous smooth chaotic process, such as the first component of the classical Lorenz equations (L63): $\{\dot{\tilde{x}} = \tau^{-1}\sigma(\tilde{y}-\tilde{x}), \dot{\tilde{y}} = \tau^{-1}(\rho \tilde{x}-\tilde{y}-\tilde{x}\tilde{z}), \dot{\tilde{z}} = \tau^{-1}(-\beta\tilde{z}+\tilde{x}\tilde{y})\}$, {\hl with an appropriate choice for the} {\hll time-scale-tuning} parameter $\tau$ {\hl as follows. With the common choice for a chaotic solution of the original ($\tau=1$) equations: $\sigma = 10$, $\rho = 28$, $\beta = 8/3$, the time scale of L63 is $\tau_{L63}=0.7$. The latter is defined by the crossover frequency in the power spectrum. Setting some other value for the {\hll time-scale-tuning} parameter we can achieve a new time scale: $\tau\tau_{L63}$. Therefore, a comparable time scale can be achieved by $\tau=\tau_{L84}/\tau_{L63}$. However, it is not $\tau\tau_{L63}$ that we want to set equal to $\tau_{L84}$. But rather, we regard the time scale of the driving comparable or approximately equal to the intrinsic time scale when L63 exerts maximal response of the driven L84 in terms of extremal behavior, measured e.g. by either the kurtosis or a high quantile of the distribution of $x$, as found in~\cite{PhysRevE.87.022822}. We can call this a kind of resonance, and define a time scale $\tau'$ such that it is unity in resonance.} Based on our finding reported in~\cite{PhysRevE.87.022822} $\tau'\approx0.4\tau$. Note that $\tau'\approx2.3>1$ when $\tau\tau_{L63}=\tau_{L84}$. Beside the resonant $\tau'=1$ scenario {\hl(M2)} we will also consider one when the driving is much faster than the model climate: $\tau'=1/4$ {\hl(M3)}. {\hll Assuming an even larger time scale separation between the resolved and unresolved processes one can apply an {\em uncorrelated} white noise (WN) driving: $\tilde{x}(t)=\xi$, $\int_{-\infty}^tdt\xi=W_t$, where $W_t$ is a Wiener process. {\hl We label the resulting model by M4.}} We use a coupling strength $A=0.025$ in case of the L63-driving, and an appropriate choice of $A$ in case of the WN-driving that gives the same variance of the driving.

\section{Results}\label{sec:results}

{\hll In this paper we focus primarily on the predictability of threshold exceedances of the first component $x$ of the various L84 models M1-4, whose symbol happens to coincide with that of our generic observable $x$. We will consider one other observable, and if it is not explicitly said, we mean to speak about the main observable $x$. Also, unless explicitly otherwise said, figures for $D$ are based on the $\mathcal{L}$-ROC curve.}

\subsection{\hll Predictability of peak-over-threshold events}

{\hll Histograms in this paper are constructed from sets of about $5\times10^6$ discrete data points each, resulting from appropriately long simulations. For the following results we simulate the autonomous and L63-driven L84 {\hl(M1-3)} using Matlab's \texttt{ode45}, which integrator chooses the time step size $h$ adaptively. This is to make use of the event-handling capability of \texttt{ode45} for the purpose of locating smooth apexes of $x(t)$. We employ the explicit order 1.5 strong scheme described in~\cite{PhysRevE.87.022822} to integrate {\hl M4} the WN-driven L84 with fixed $h=0.01$.}

\subsubsection[Makeup]{{\hll Dependence on the makeup of the precursory structure and on intrinsic properties the model}}\label{sec:observable_choice}

\begin{figure*} %[t!]
    \begin{center}
	  \includegraphics[width=\linewidth]{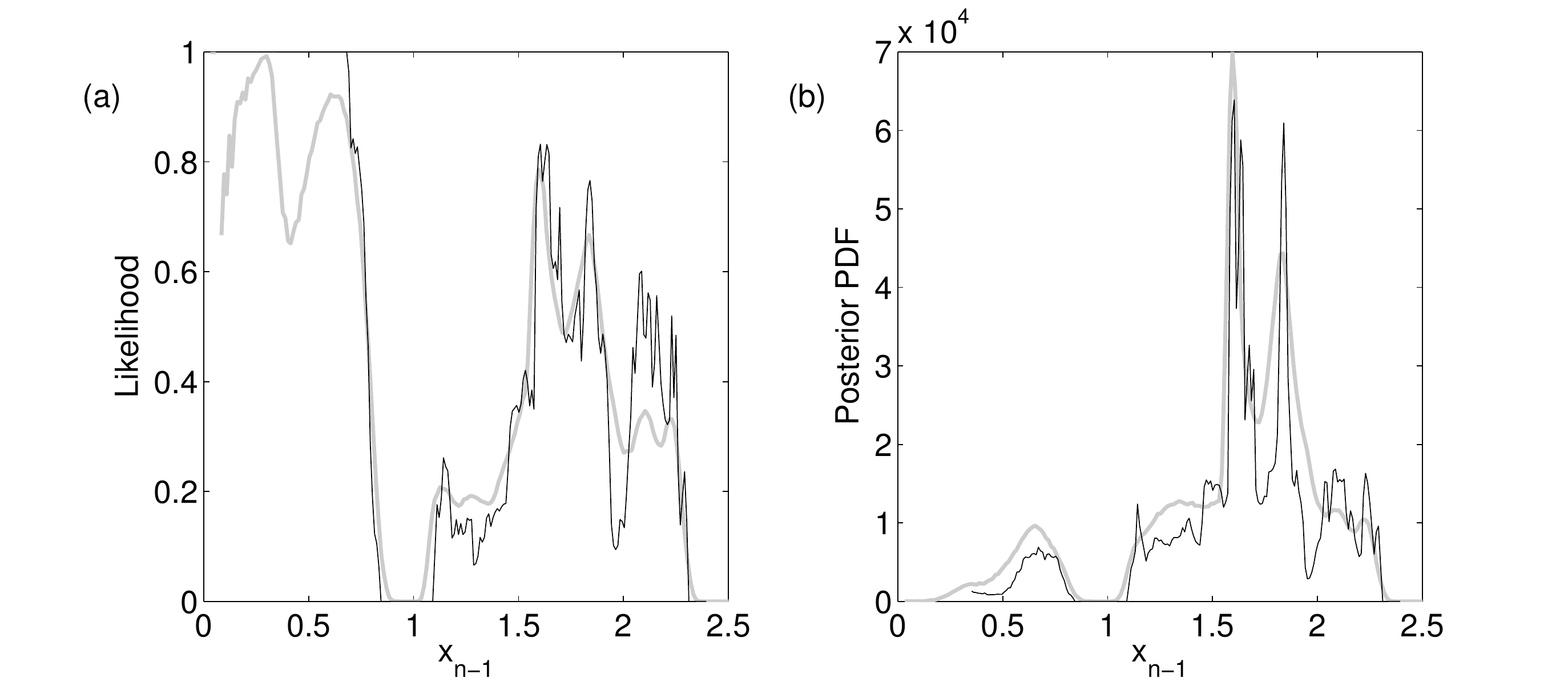}
        \caption{\label{fig:Pp_L_x} Distributions of the likelihood $\mathcal{L}$ and posterior probability density $\mathcal{P}$ when $\mathbf{x}_n=x_{n-1}$ {\hl and $x_*=1.8$}. The solid thin black line and thick gray line correspond, respectively, to {\hl models M1 and M3 defined in Sec. \ref{sec:model}. Corresponding ROC curves in case of M1 are shown in Fig. \ref{fig:ROC_Pp_L_x}.} 
        }
    \end{center}
\end{figure*}

The most simple case of a discrete-time precursory structure is that of the previous peak value of the observable whose threshold exceedances are to be predicted: $\mathbf{x}_n=x_{n-1}$. The two panels of Fig. \ref{fig:Pp_L_x} show the posterior PDF and the likelihood function, respectively, represented by adequate histograms. The cases of the autonomous, i.e., undriven, L84 {\hl (M1)}, and the L63-driven L84 {\hl(M3)} are shown in one diagram side-by-side. As expected, the driving smooths out features of both distributions {\hl seen in  Fig. \ref{fig:Pp_L_x}}, however, contrary to expectations: e.g. the likelihood (b) can be even enhanced by driving (see for example $1.2<x_{n-1}<1.4$). Furthermore we point out that the relationship given by Eq. (\ref{eq:likelihood}) is manifested in the more broad structures of the distribution of the likelihood as compared to that of the posterior probability density. {\hll This is so because $\mathcal{P}(\mathbf{x})$ tends to be peaked where $p(\mathbf{x})$ is peaked.} As already mentioned in Sec. \ref{sec:pred_data_driv}, this broadening ought to be reflected in the relative positions of the respective ROC curves. In fact the pair of ROC curves in Fig. \ref{fig:ROC_Pp_L_x} belong to the present prediction scenario considering {\hl M1}.

\begin{figure*} %[t!]
    \begin{center}
        \begin{tabular}{c}
            \includegraphics[width=\linewidth]{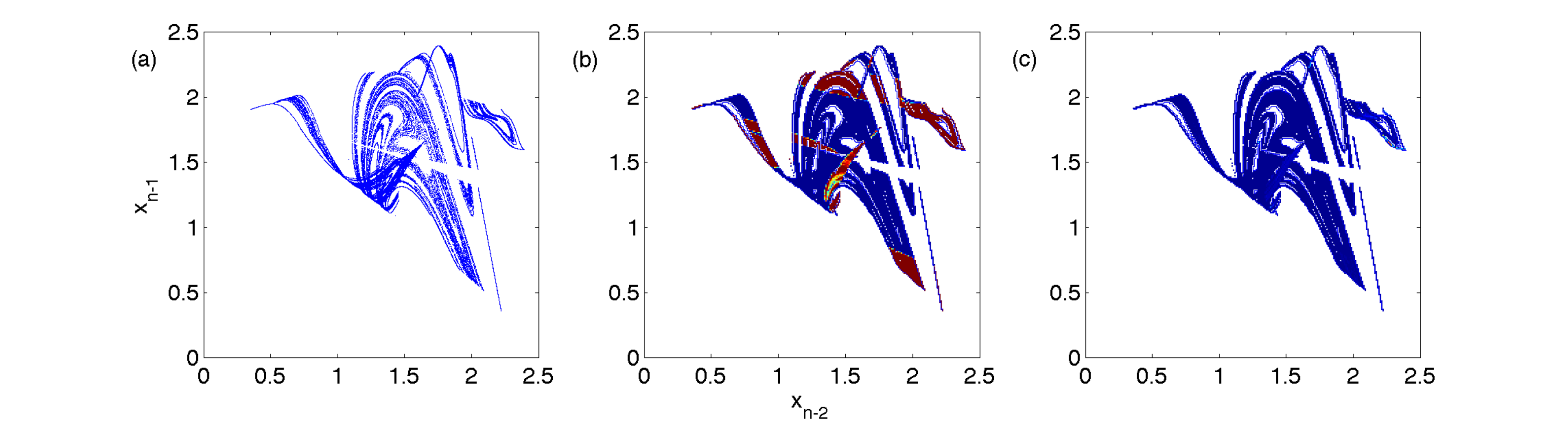} \\
            \includegraphics[width=\linewidth]{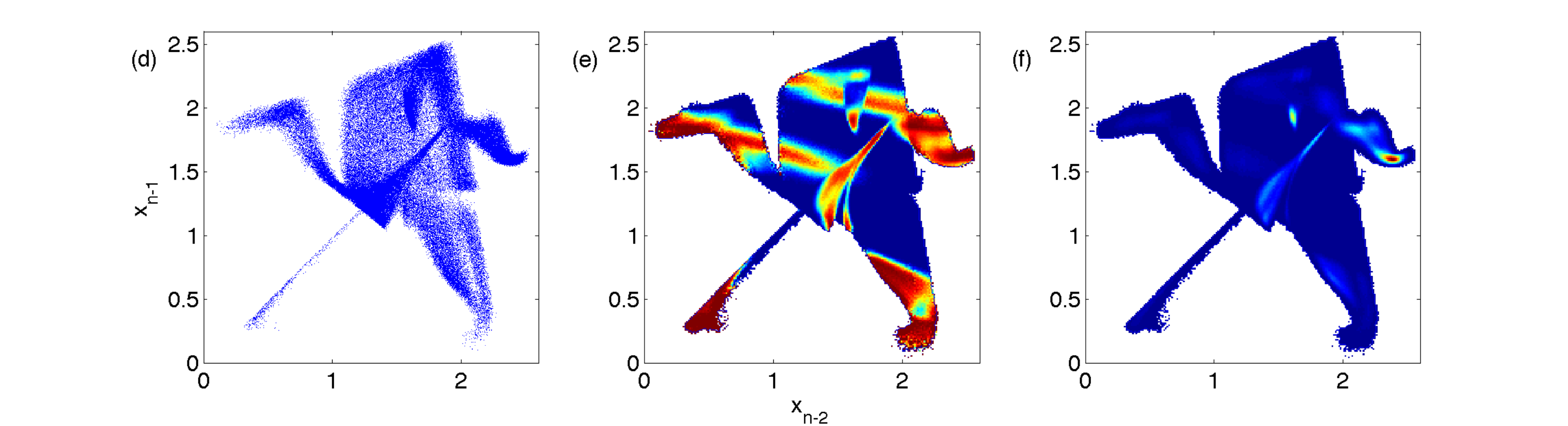} \\
            \includegraphics[width=\linewidth]{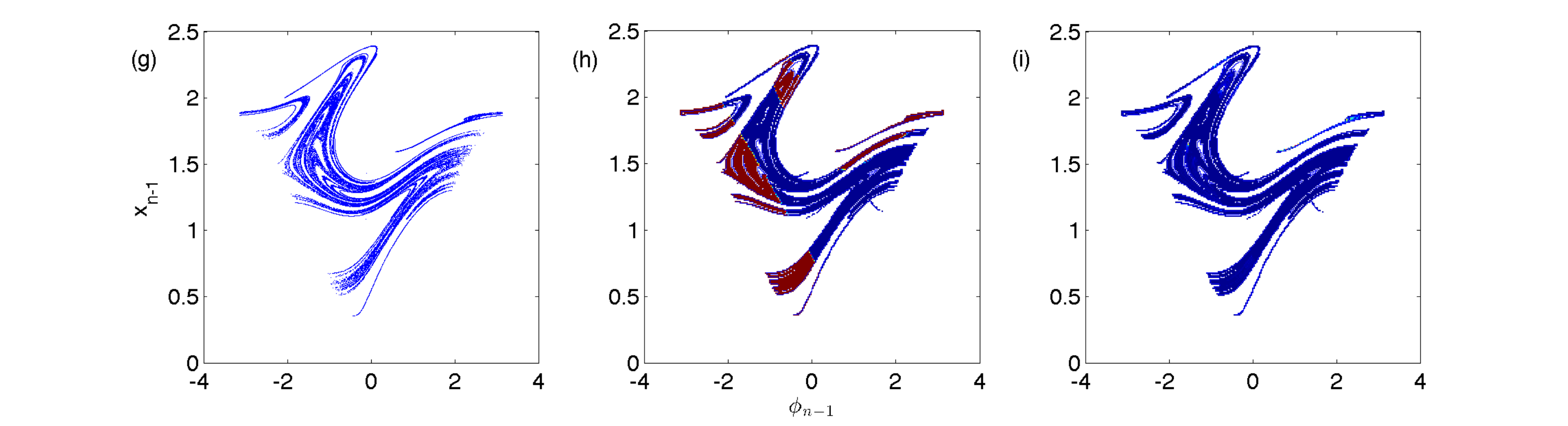} \\
            \includegraphics[width=\linewidth]{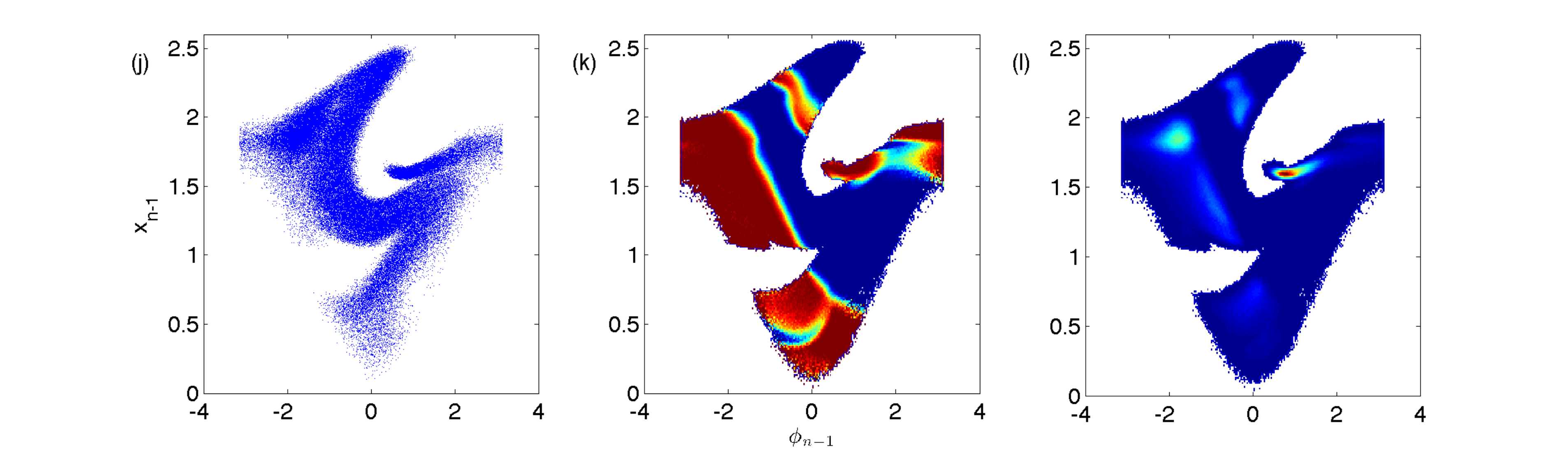}
        \end{tabular}
        \caption{\label{fig:Pp_L_xx_xphi} Scatter plots (left) and distributions of the likelihood $\mathcal{L}$ (middle) and posterior probability density $\mathcal{P}$ (right). In each diagram the color scale ranges from dark blue for 0 to dark red for the maximal density value, which is unity for $\mathcal{L}$ for the presented scenarios, but various different values for $\mathcal{P}$. {\hll We used $x_*=1.8$. The first and third (second and fourth) row concern M1 (M3). The following correspondence between the labeled panels above and scenarios labeled by boxed numbers in Table \ref{tab:distance_values_x} and Fig. \ref{fig:ROC_all} stand, respectively: (b) 7 (c) 8 (e) 5 (f) 6 (h) 11 (i) 12 (k) 9 (l) 10.} }
    \end{center}
\end{figure*}

It is an intuitive expectation that the predictability can be improved by relying on more information by means of extending the precursory structure. The most simple step in this direction -- in the realistic context when only one variable is available or practical to observe -- is that beside the previous data point we monitor also the one just before that, i.e., $\mathbf{x}_n=(x_{n-1},x_{n-2})$. The distributions for the same scenarios as considered before in Fig. \ref{fig:Pp_L_x} are displayed in Fig. \ref{fig:Pp_L_xx_xphi}. The bivariate distributions are visualized by color plots; and beside the distributions, on the left we also display scatter plots of data points (but for a better visibility of features we plot fewer points than those that the histograms are constructed from). In the case of the autonomous L84 {\hl(M1)} the scatter plot {\hl[Fig. \ref{fig:Pp_L_xx_xphi} (a)]} reveals a fractal pattern {\hll with a distinctive filamentary structure}. This could lead one to think that there is a one-to-one or unique relationship between subsequent pairs of $(x_{n-1},x_{n-2})$, which is also called a mapping or map~\cite{Tel_n_Gruiz:2006}. This can be confirmed by looking at the distribution of the likelihood {\hl[Fig. \ref{fig:Pp_L_xx_xphi} (b)]}, which takes on the maximum value of unity wherever the scatter plot exhibits fractality. -- Because of the uniqueness, we have a deterministic rule to predict the next value, and so we can tell with certainty whether it will exceed the threshold. In regions where a lack of clear fractality is observed, e.g. around $(x_{n-1},x_{n-2})=(1.4,1.4)$, $\mathcal{L}<1$ consistently. The exhibited pattern of the scatter plot can give the intuition that the lack of uniqueness is a result of `looking at' a curved surface living in 3D `from a poor angle' so that the 2D view of some parts of the surface is obstructed by other parts of it. In other words, the surface looks folded. In fact, Takens' embedding theorem~\cite{Takens:1981} states that an attractor of Hausdorff dimension $D_0$ can (always) be embedded by $M>2D_0$ number of delay variables. For us this means an unfolded appearance. In our case $D_0\approx 1.6$~\cite{PhysRevE.87.022822}, and so for uniqueness we need maximum $M=4$. This does not `encourage' us that we can have an unobstructed 2D view, although neither does it say that we cannot have. In fact, in our case we can have such a view, to be described next.

Let us bear in mind that the discrete $x_n$ data belong to apexes of the continuous $x(t)$. In these points $\dot{x}=0$. We can use this fact in conjunction with the first component of the equations of L84 (\ref{eq:L84}) to determine that the apexes are situated on the surface: $y^2+z^2=-x/4+aF$. This can be viewed as a Poincar\'e surface of intersection that defines a slice of the attractor -- the Poincar\'e section~\cite{Tel_n_Gruiz:2006}. For any fixed $x$ we recognize the equation of a circle. That is, the surface itself is locally conical, which approximation applies well to the chaotic attractor with $F=8$ extending between about [-0.5,2.5] wrt. $x$, as seen in Fig. \ref{fig:time_history}. Such a surface can be rectified on the plane spanned by the azimuthal angle 

\begin{equation}\label{eq:phi}
  \phi=\arctan(y/z),
\end{equation} 
%$$\phi=\arctan(y/z),$$ 
%
periodic in e.g. $[0,2\pi]$, and $x$. Therefore, there exists a unique mapping between subsequent pairs of $(x_n,\phi_n)$. This allows for an unfolded view of the Poincar\'e section, and so for the prediction of the next apex with certainty. Accordingly, as seen in Fig. \ref{fig:Pp_L_xx_xphi} {\hl(g)}, the scatter plot exhibits fractality everywhere, and $\mathcal{L}=1$ (or 0) also everywhere {\hl[Fig. \ref{fig:Pp_L_xx_xphi} (h)]}. This certainty is compromised in the numerics only by the effect of coarse-graining, when $\{{\mathcal{L}}_b\}$ may be less than unity due to the finite data set size.

By introducing a driving as defined in Sec. \ref{sec:model}, the dimensionality of the problem increases. Therefore, the same precursory structure of only two variables is inevitably insufficient for predictions with certainty. {\hl The example of M3 shows that the} scatter plot becomes area-filling {\hl[Fig. \ref{fig:Pp_L_xx_xphi} (d) and (j)]}, and we will have distributions of the likelihood that take on all values between 0 and 1 {\hl[Fig. \ref{fig:Pp_L_xx_xphi} (e) and (k)]}. However, as long as the forcing strength is moderate, the new features tend to develop through a smearing of the original ones. Evidently, this applies partially in our case.

\begin{figure} [t!]
    \begin{center}
	  \includegraphics[width=\linewidth]{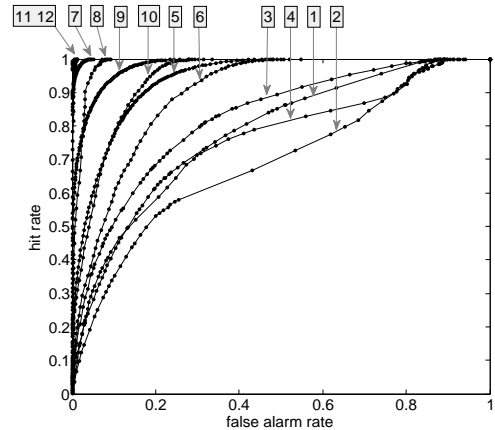} 
        \caption{\label{fig:ROC_all} $\mathcal{L}$- and $\mathcal{P}$-ROC curves for scenarios specified in Table \ref{tab:distance_values_x} marked with the corresponding numbers framed in boxes.}
    \end{center}
\end{figure}

ROC curves that derive from the distributions seen in Figs. \ref{fig:Pp_L_x} and \ref{fig:Pp_L_xx_xphi} are shown in Fig. \ref{fig:ROC_all}. $\mathcal{L}$-ROC curves are always above the $\mathcal{P}$-ROC curves, and the corresponding ROC curves for undriven and driven versions of L84 have the same relation consistently. In particular, for the scenario of the undriven L84 when $\mathbf{x}_n=(x_{n-1},\phi_{n-1})$, the ROC curves approach very near the ideal corner of certainty. 

For the various scenarios considered the distances $D$ (\ref{eq:distance}) from the corner, as a summary measure, is provided in Table \ref{tab:distance_values_x}. Besides the scenarios treated in Fig. \ref{fig:ROC_all}, data is provided in the table also for several other scenarios as follows. First off, we used white noise driving too. Unexpectedly, the predictability with a single delay variable, $\mathbf{x}_n=x_{n-1}$, is better for this driven case than the undriven one. This is unchanged even if we apply a smoothing to the time series. Note that $x$ of the WN-driven L84 {\hl(M4)} is a red noise-like nonsmooth process. An improvement of predictability by smoothing is achieved only with larger precursory structures {\hl({\hll compare the values between round brackets with the numbers just above each} in the said table)}. {\hl The effect of improved predictability having introduced a driving} could be due to a {\em stabilization} of the trajectory by noise. However, instead of $x$, considering another observable, namely, the total cyclonic activity in the model, 

$$r=\sqrt{y^2+z^2},$$ 
{\hl we do not observe the same effect}; see Table \ref{tab:distance_values_r}, and note that the process of $r$, contrary to that of $x$, is smooth. Therefore, the more likely cause of the unexpected effect is that the geometry of the attractor is altered by noise in a favorable manner, when the Poincar\'e section `looks' {\em less folded} `in view of $x$' (but not $r$). 

Beside a fast ($\tau'=1/4$) L63-driving {\hl(M3)} we also considered a slower one with matching time scales of driving and model climate, i.e., $\tau'=1$ {\hl(M2)}. The rationale for this is that we expect that the delay variables, with delay times determined dominantly by the main system, would be able to pick up more information on the driving of a longer decorrelation time. However, while this mechanism should be at work, an improvement of predictability is not registered, but on the contrary. {\hl This is so} also when using one more delay variable: $\mathbf{x}_n=(x_{n-1},x_{n-2},x_{n-3})$, or {\hl when} considering observable $r$. The likely cause of this is that, as a counter-effect, the trajectory is destabilized by the driving more so with $\tau'=1$ than 1/4. 

Beside -- but not independent of -- the issue of the {\em choice} of the observables {\hl to make up the precursory structure}, we can make an interesting observation to do with the {\em size} of the precursory structure too. Extending the two delay variables with a third one (all of the same type) did improve the predictability. However, it was still not as good as with the shorter precursory structure involving the azimuthal angle $\phi$. This is consistent with Takens' embedding theorem, as mentioned earlier.

\begin{table*}
\caption{Summary measure for the ROC-statistics: the `distance' $D$ (\ref{eq:distance}) from the ideal case of all events correctly predicted without making any false alarms. {\hl The models M1-4 assessed are defined in Sec. \ref{sec:model}. The threshold is the same in all cases, that is, $x_*=1.8$. Histograms were constructed with $B^{1/2}=200$ bins % CONFIRMED BY INFORMATION IN MY PAPER NOTES 
in all dimensions in the respective ranges where data points are present; the `resolution' that the $B^{1/2}=200$ bins and the $N\approx5\times10^6$ data points give is depicted in the middle and right columns of Fig. \ref{fig:Pp_L_xx_xphi}.} The number of significant digits were determined based on only two independent realizations. The figures in round brackets were obtained by a smoothing of the nonsmooth white noise-driven time series over a moving window of width of a nondimensional time unit. The numbers framed by boxes indicate the correspondence with results shown in Fig. \ref{fig:ROC_all}. 
}\label{tab:distance_values_x}
  \begin{center}
    \begin{tabularx}{1\textwidth}{c *8{>{\Centering}Y}}     %{c|cc|cc|cc|cc}       {c *{8}{Y}}
      \toprule
      %\multirow{2}{*}{ \vtop{\hbox{\strut Precursory}\hbox{\strut structure}}} & \multicolumn{2}{c}{Undriven L84} & \multicolumn{2}{c}{L63-driven *, $\tau'=1$} & \multicolumn{2}{c}{L63-driven *, $\tau'=1/4$} & \multicolumn{2}{c}{WN-driven *} \tabularnewline
      \multirow{2}{*}{ \vtop{\hbox{\strut Precursory}\hbox{\strut structure}}} & \multicolumn{2}{c}{M1} & \multicolumn{2}{c}{M2} & \multicolumn{2}{c}{M3} & \multicolumn{2}{c}{M4} \tabularnewline
      \cmidrule{2-9} & $\mathcal{P}$ & $\mathcal{L}$ & $\mathcal{P}$ & $\mathcal{L}$ & $\mathcal{P}$ & $\mathcal{L}$ & $\mathcal{P}$ & $\mathcal{L}$ \tabularnewline \midrule
      %$x_{n-1}$ & 0.41 & 0.36 & 0.48 & 0.41 & 0.49 & 0.41 & 0.39 & 0.36 \tabularnewline \midrule
      $x_{n-1}$ & 0.414 \fbox{4} & 0.359 \fbox{3} & 0.480 & 0.409 & 0.490 \fbox{2} & 0.410 \fbox{1} & 0.390 (0.374) & 0.357 (0.358) \tabularnewline \midrule
      $(x_{n-1},x_{n-2})$ & 0.062 \fbox{8} & 0.027 \fbox{7} & 0.278 & 0.217 & 0.265 \fbox{6} & 0.197 \fbox{5} & 0.527 (0.312) & 0.320 (0.226) \tabularnewline \midrule
      $(x_{n-1},\phi_{n-1})$ & 0.0108 \fbox{12} & 0.0054 \fbox{11} & 0.1803 & 0.1116 & 0.1840 \fbox{10} & 0.1114 \fbox{9} & 0.2930 (0.206)  & 0.2587 (0.156)  \tabularnewline \midrule
      $(x_{n-1},x_{n-2},x_{n-3})$ & 0.0177 & 0.0087 & 0.2456 & 0.1434 & 0.2257 & 0.1350 & 0.4917 (0.323) & 0.2828 (0.189) \tabularnewline
      \bottomrule
    \end{tabularx}
  \end{center}
\end{table*}

\begin{table*}
\caption{Same as in Table \ref{tab:distance_values_x} but with observable $r$ of cyclonic activity, {\hl and $r_*=1.8$.} % CONFIRMED BY RERUNNING SIMULATION WITH ProbSI11.m loading mainL84turn1.mat setting xth=1.8 (which is what is denoted by r_* here), retaining the right def' of 'XE' and recovering gop=0.0069 /o/  
The number of significant digits is taken to be the same as in case of observable $x$, i.e., not based on a number of independent realizations.
}\label{tab:distance_values_r}
  \begin{center}
    \begin{tabularx}{1\textwidth}{c *8{>{\Centering}Y}}     %{c|cc|cc|cc|cc}       {c *{8}{Y}}
      \toprule
      %\multirow{2}{*}{ \vtop{\hbox{\strut Precursory}\hbox{\strut structure}}} & \multicolumn{2}{c}{Undriven L84} & \multicolumn{2}{c}{L63-driven *, $\tau'=1$} & \multicolumn{2}{c}{L63-driven *, $\tau'=1/4$} & \multicolumn{2}{c}{WN-driven *} \tabularnewline
      \multirow{2}{*}{ \vtop{\hbox{\strut Precursory}\hbox{\strut structure}}} & \multicolumn{2}{c}{M1} & \multicolumn{2}{c}{M2} & \multicolumn{2}{c}{M3} & \multicolumn{2}{c}{M4} \tabularnewline
      \cmidrule{2-9} & $\mathcal{P}$ & $\mathcal{L}$ & $\mathcal{P}$ & $\mathcal{L}$ & $\mathcal{P}$ & $\mathcal{L}$ & $\mathcal{P}$ & $\mathcal{L}$ \tabularnewline \midrule
      %$r_{n-1}$ & 0.47 & 0.33 & 0.51 & 0.37 & 0.50 & 0.35 & 0.60 & 0.42 \tabularnewline \midrule
      $r_{n-1}$ & 0.475 & 0.329 & 0.513 & 0.368 & 0.503 & 0.354 & 0.599 & 0.420 \tabularnewline \midrule
      $(r_{n-1},r_{n-2})$ & 0.054 & 0.025 & 0.192 & 0.123 & 0.171 & 0.098 & 0.288 & 0.181  \tabularnewline \midrule
      $(r_{n-1},r_{n-2},r_{n-3})$ & 0.0150 & 0.0069 & 0.1667 & 0.0802 & 0.1433 & 0.0660 & 0.2829 & 0.1440  \tabularnewline
      \bottomrule
    \end{tabularx}
  \end{center}
\end{table*}

\subsubsection{\hll Dependence on the event-magnitude}

{\hll

We evaluate now the dependence of predictability on the event-magnitude to achieve our objective (i.b). That is, we construct $D(x_*)$, with computations for an array of sample points of $x_*$. The numerical value of $D(x_*)$ does vary with the bin size $\Delta x$. This variation has a single minimum in all cases checked (and so we assume that this is always the case), but these are different $\Delta x_{opt}$ values for different values of $x_*$. We intend to construct the $D(x_*)$ which is optimized for all values of $x_*$, and this is what we regard as the `dependence of predictability on the event-magnitude'. This result is achieved by constructing the unoptimized $D(x_*)$ for a range of $\Delta x$ (or $B$) values, and plot these curves in a single diagram. The range of $\Delta x$ values should include the optimal values belonging to all $x_*$'s. Then, the lower envelope of these curves will represent the optimized $D(x_*)$. In Fig. \ref{fig:d_vs_ths_pot} this construction is shown for M1, M3, M4, side-by-side, considering the precursory structures  $\mathbf{x}_n=(x_{n-1},x_{n-2})$ and $\mathbf{x}_n=(x_{n-1},\phi_{n-1})$, $\phi$ being defined by (\ref{eq:phi}). For each model we see a decreasing trend of $D(x_*)$, but only for the largest values of $x_*$ (from about 1.8), and only on coarse scales of $x_*$, i.e., on smaller scales of $x_*$ the variation of $D(x_*)$ can be nonmonotonic\footnote{\hll A blowup approaching the largest value of $x$, and so $x_*$, is due to the undersampling of the probability distributions, and so it is to be disregarded. The undersampling does not show up for M1 because the tail of the process PDF $p_x(x_*)$, to be defined shortly below, does not decay slowly like in the other models.}. In effect we have assessed the monotonicity of the magnitude-dependence $D(x_*)$ depending on other factors, the precursory structure and also the model, and we have found a rather robust behavior. 

\begin{figure*} %[t!]
    \begin{center}
        \includegraphics[width=\linewidth]{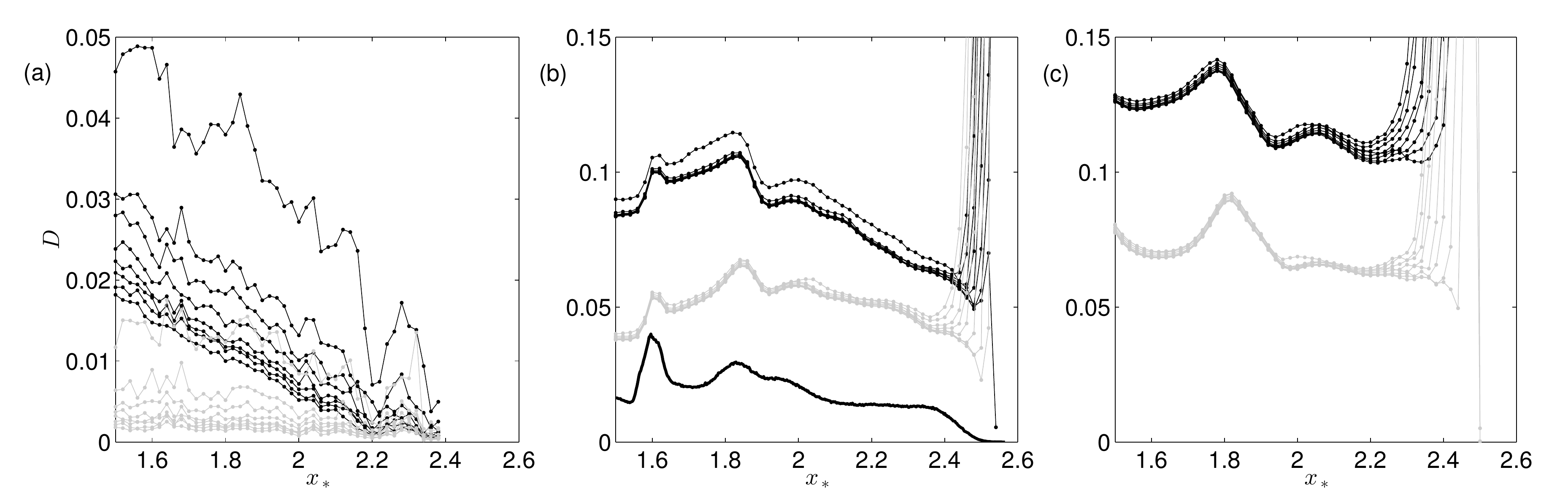} % 0.6\linewidth works but the full \linewidth of 0.7\linewidth don't. Or rather: this figure shows up, but some of the later ones do not!!! 0.9 is OK when clearing the bf-highlights.
        \caption{\label{fig:d_vs_ths_pot} \hll Event magnitude-dependence of predictability of POT events. The thin black and gray lines depict the dependence using the precursory structure $(x_{n-1},x_{n-2})$ and $(x_{n-1},\phi_{n-1})$, respectively, for models (a) M1, (b) M3 (result qualitatively representative of those for M2), (c) M4. In each scenario a bundle of curves belong to $B^{1/2}=50j$, $j=1,\dots,8$ number of bins in one dimension. The lower envelope of the bundle approximates the optimized $D(x_*)$. In panel (a) the optimal $B^{1/2}$ is larger than 400 (and checked to be larger than even 1000), but we believe that the same qualitative behavior persist up to the optimum. Dot markers on the curves mark out sample values of $x_*$. In panel (b) the (rescaled) PDF of $x$ is included for reference, depicted by the lowermost thick black line.}
    \end{center}
\end{figure*}

Next we carry out a thought experiment raising continuously a horizontal line in the scatter plots in the left column of Fig. \ref{fig:Pp_L_xx_xphi}. We monitor for what $x_{n-1}$ levels do new features enter into- or features present exit from what is above the line. The measure of points above the line is in fact given by the denominator in (\ref{eq:hit_rate}) as the integral of the posterior probability density, which is nothing but $\mathbb{P}_{\chi}(\chi = 1)$, because $\int_{\mathbb{R^M}}dV_{\mathbf{x}} p_{\mathbf{x}| \chi}(\mathbf{x} , \chi = 1)=1$. Considering the meaning of the event variable (\ref{eq:chi}), $\mathbb{P}_{\chi}(\chi = 1)$ as a function of the threshold $x_*$ only is equivalent with the complementary distribution function belonging to the process PDF $p_x(x_*)$ of one variable $x_*$ pertaining to the observable of concern $x$. Therefore, what we monitor in effect is the points or levels $x_*$ of discontinuities or fast changes of $p_x(x_*)$. The levels of the major discontinuities are discernible in Fig. \ref{fig:Pp_L_xx_xphi}. Note that in the unperturbed L84 because of the quadratic tangency of the filaments to horizontal lines, for increasing $x_*$, before a discontinuous drop of the density $p_x(x_*)$, it actually increases. In the driven L84 there are no discontinuities, they are `washed out', but, as the lowermost curve in panel (b) of Fig. \ref{fig:d_vs_ths_pot} shows, the density $p_x(x_*)$ features strong nonmonotonicities at the levels of interest. We observe that these levels coincide well with the locations of `humps' of $D(x_*)$ in all cases displayed in Fig. \ref{fig:d_vs_ths_pot} (a) and (b). This suggests that the behavior of $D(x_*)$ is controlled at least in part by $p_x(x_*)$. This raises the question whether the decay of $p_x(x_*)$ is responsible for the decreasing nature of $D(x_*)$, at least in the present situation. We will revisit this question in Sec. \ref{sec:summary}.

}

\subsection{\hll Predictability of threshold-exceedance-in-an-interval events}

{\hll Here we attend to our objective (iii).}

\subsubsection{{\hll Dependence on the} prediction lead time}\label{sec:lead_time}

% HAVING MADE THE FOOTNOTE THAT SEC  \ref{sec:observable_choice} CAN BE SKIPPED AND REREAD THE FOLLOWING TWO PARA, I THOUGHT THAT THAT SUBSEC CANNOT BE SKIPPED. BUT ON RECONSIDERATION I THINK AGAIN THAT IT CAN BE. 

Let us emphasize that {\hl when the precursory space can embed the attractor {\hll -- defining an ideal precursory structure --}} the effect of the destabilization of trajectories, mentioned {\hll in Sec. \ref{sec:observable_choice}}, can influence data-driven predictability only in the practical sense of having a {\em finite} trajectory length, i.e., finite data set size. Because of the latter, a coarse-graining is inevitable in constructing the histograms. When establishing a correspondence of DDP of POT events with on-demand MDP in terms of the Poincar\'e mapping (not the original flow), the coarse-graining can be translated into terms of errors in measuring initial conditions/precursory observables\footnote{For this point it does not matter whether the Poincar\'e mapping can be constructed analytically to facilitate the on-demand MDP of POT events. In fact, it is not possible in general even for the most simple chaotic flows. In that case only archival MDP of POT events is possible, whose skill, nevertheless, should be the same as that of the hypothetical on-demand MDP.} of size bounded from above by the histogram bin size. In case of a chaotic trajectory the error in tracing the trajectory {\hl forward in time} grows exponentially fast (at least while the error is still small). {\hl In MDP this manifests in the spreading out of the ensemble. Fixing the ensemble size, the evaluation of the likelihood of an event will have a larger statistical error the more spread-out the ensemble is. The latter would typically correspond to a longer prediction lead time. The larger errors in estimating the likelihood should clearly precipitate in a deterioration of the overall prediction skill. Because of the correspondence, this deterioration carries over to DDP. Note that when the attractor cannot be embedded in the precursory space, the instabilities of trajectories take effect also if $N\rightarrow\infty$, because some initial conditions are randomly initialized.}

In the situation with an ideal precursory structure as discussed {\hll in Sec \ref{sec:observable_choice}}, the likelihood was evaluated to be nil or unity, or that with a very good approximation, because the prediction lead time was {\hl in fact} limited by the typical time scale of the system, given that from one apex we intended to predict the next one. We could evaluate the dependence of predictability on the lead time by looking further than the next apex to predict. However, instead of this exercise we prefer to map out the predictability as a continuous function of the prediction lead time instead of its discrete advances. {\hl Our preference is partly due to the fact that the discrete advances} are not known `apriori', i.e., before integrating the system. That is, next we examine the predictability of not POT but TEI events.

In practice $T$, $\Delta T$, and $t$ can take values that are integer multiples of the trajectory sampling time, which latter is chosen small anyway in order to secure good accuracy of tracing out trajectories by numerical integration of (\ref{eq:L84}). In fact, for this exercise we use the classical Runge-Kutta algorithm/stochastic integrator mentioned above in case of {\hl M1-3/M4} %the undriven/WN-driven L84 
with fixed $h=0.01$, and we save the state in every 5th time step (to make sure that there will never be two trajectory points subsequent in time in one bin); furthermore we apply $\Delta T=2\times5\times h$. As for the precursory structure we will take the triplet of the system state variables $(x,y,z)$.  Therefore, the threshold $x_*$, the interval length $\Delta T$, and the dynamics itself, determine an {\em event volume} in phase space. We then generate data points for the tri-variate histogram $\{{\mathcal{P}}_b\}$ by identifying trajectory sample points in the event volume and trace them backward in time by $T$. {\hl We do this not on-the-fly during the simulation, but as a postprocessing of the pregenerated long time series data produced by numerical integration. For now} we choose a bin size {\hl rather} arbitrarily, and by `predictability' {\hll now} we mean {\hll predictability} conditional to the fixed bin size, not the best possible predictability -- given a fixed data set -- as a result of some optimization.

\begin{figure} %[t!]
    \begin{center}
	  \includegraphics[width=\linewidth]{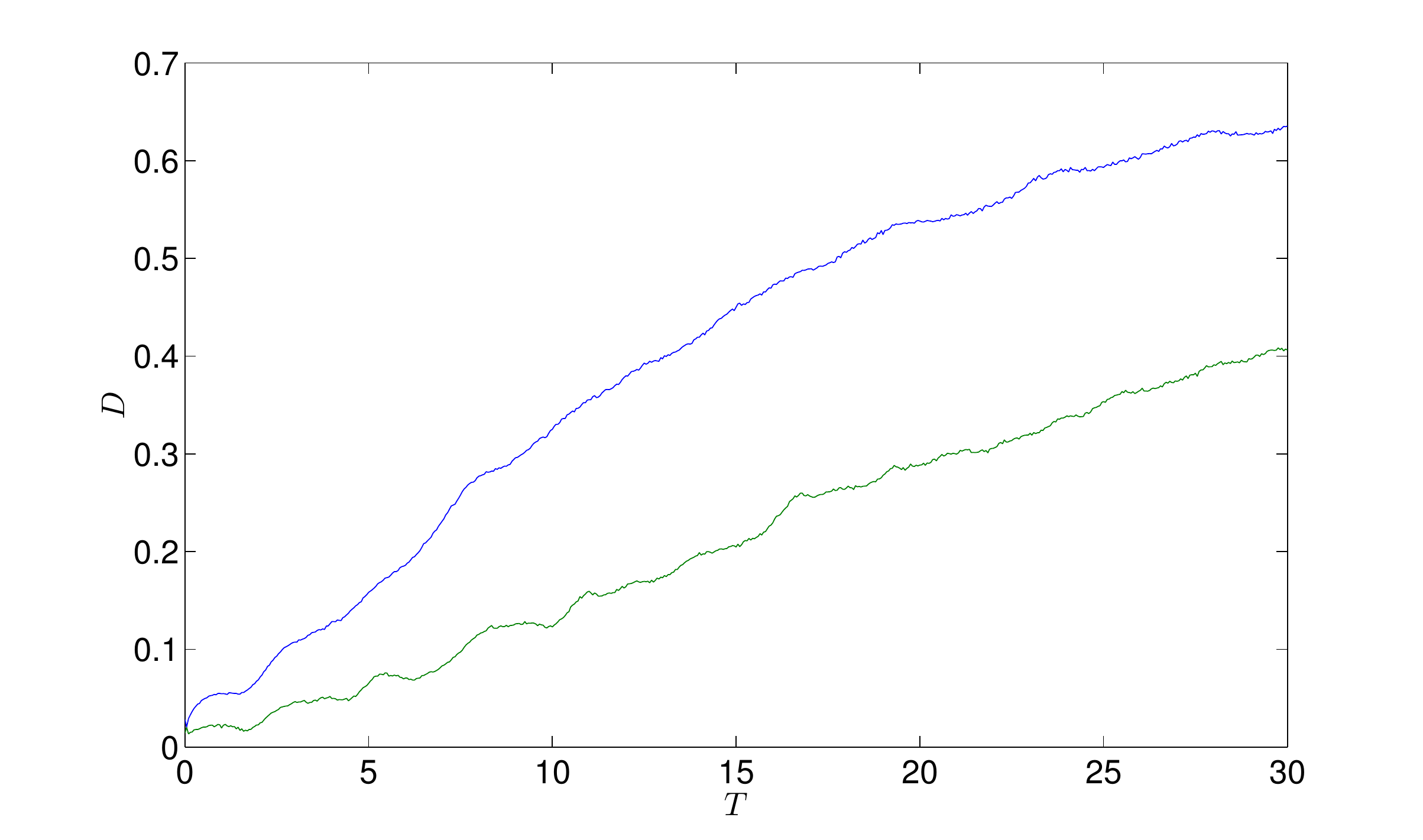} 
        \caption{\label{fig:d_vs_plt} Predictability as a function of the prediction lead time. The curve on top (blue) is obtained for {\hl M4}, %the WN-driven L84, 
        and the other one (green) for {\hl M1}. %the undriven L84. 
        Histograms were constructed with $B^{1/3}=40$ bins in all dimensions in the respective ranges where data points are present.}
    \end{center}
\end{figure}

The results of evaluating the predictability in terms of the distance $D$ for {\hl M1 and M4} %the undriven and the WN-driven L84 
are displayed in Fig. \ref{fig:d_vs_plt}. We can make a number of observations. First, the driven system is less predictable, as expected. Second, the predictability is declining in both cases with increasing prediction lead time, as it should, given the chaotic dynamics and the finite data set size. We have checked that beyond the range of $T$ shown the curves approach the $\sqrt{2}/2$ asymptote, belonging to a straight diagonal ROC curve, meaning no prediction skill at all. This is also expected. Third, on shorter scales $D(T)$ is {\em not monotonic} in either case, unlike in the well-known case of an auto-regressive AR(1) process of order one, studied regarding data-driven predictability by Hallerberg and Kantz~\cite{npg-15-321-2008}. In the case of L84 the deterministic or autonomous part of its equations results in a chaotic dynamics which is much more complex than the linear {\hl deterministic} term of the AR(1). In particular, the {\hl deterministic} term of AR(1) need to have a stabilizing effect on the {\hl trajectories} in order to have a bounded dynamics, while, although on the compact chaotic attractor of L84 trajectories are bounded, they are unstable in a long-term average sense measured by a positive Lyapunov exponent. This instability can also deteriorate predictability, beside a stochastic part if any. On short-terms, however, the deterministic trajectory can experience stable periods, which periods are associated with the {\em return of skill} admitted by the plateaus, or negative slopes even, of $D(T)$. 

\subsection{\hll Dependence on the event-magnitude}\label{sec:threshold}

In comparison with the AR(1) process, a further matter of interest is the dependence of predictability on the threshold level. The counterintuitive finding in case of AR(1) was reported by Hallerberg and Kantz~\cite{npg-15-321-2008}, namely, that stronger extremes -- indifferently to the distribution that the process realizes -- are more predictable. The obvious question to ask is, then, whether the latter holds also in case of processes with a more complex deterministic dynamics subjected {\hl(or not)} to stochastic forcing.

\begin{figure}  [t!]
    \begin{center}
        \begin{tabular}{c}%{cc}
            \includegraphics[width=\linewidth]{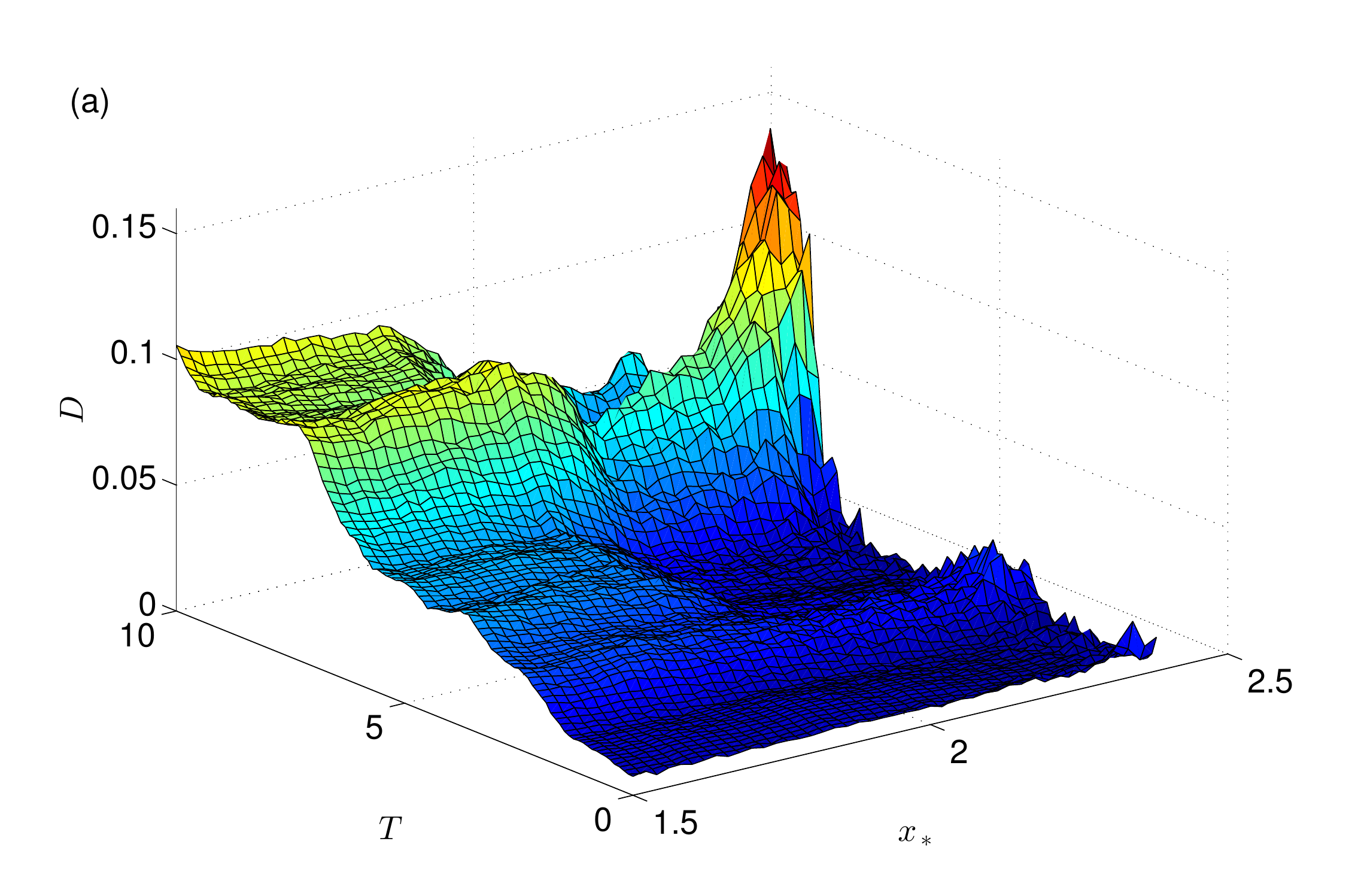} \\% & 
            \includegraphics[width=\linewidth]{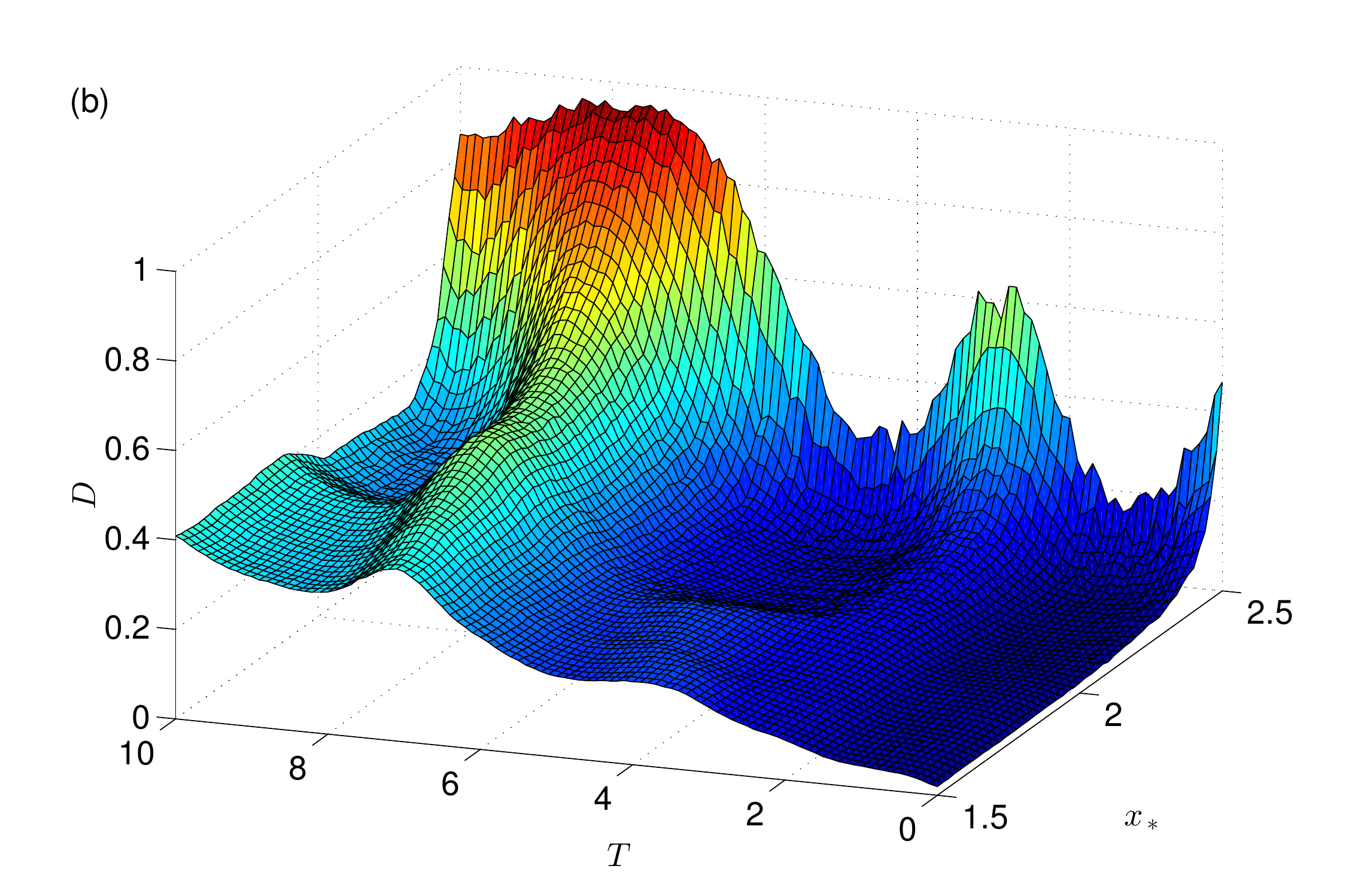} %& 
        \end{tabular}
        \caption{\label{fig:ROC_vs_ftle} Predictability in terms of $D$ %(left) and $\langle\lambda^{(T)}\rangle$ (right) 
        as a function of the prediction lead time $T$ and threshold level $x_*$. Results are shown for {\hl models M1 and M4 %the undriven (a) and the WN-driven L84 (c). 
        in panels (a) and (b), respectively.} Histograms %in association with (a) and (c) 
        were constructed with $B^{1/3}=100$ bins in all dimensions in the respective ranges where data points are present. Notice the different ranges of $D$ shown. 
        }
    \end{center}
\end{figure}

Figure \ref{fig:ROC_vs_ftle} (a) and (b) show the predictability as a function of the prediction lead time as well as the threshold level for {\hl M1 and M4}, %the cases of the undriven and WN-driven L84, 
respectively. Firstly, the {nonmonotonic} nature of $D(T)$ is prevalent on any fixed threshold level. Secondly, we observe that while for some fixed prediction lead times stronger events are more predictable, i.e., $D(x_*)$ is a decreasing function, it is just the opposite for some other $T$'s. That is, the above statement for AR(1) does not seem to hold in general for more complex dynamical systems. 

\begin{figure*}  [t!]
    \begin{center}
        \begin{tabular}{cc}
            \includegraphics[width=0.5\linewidth]{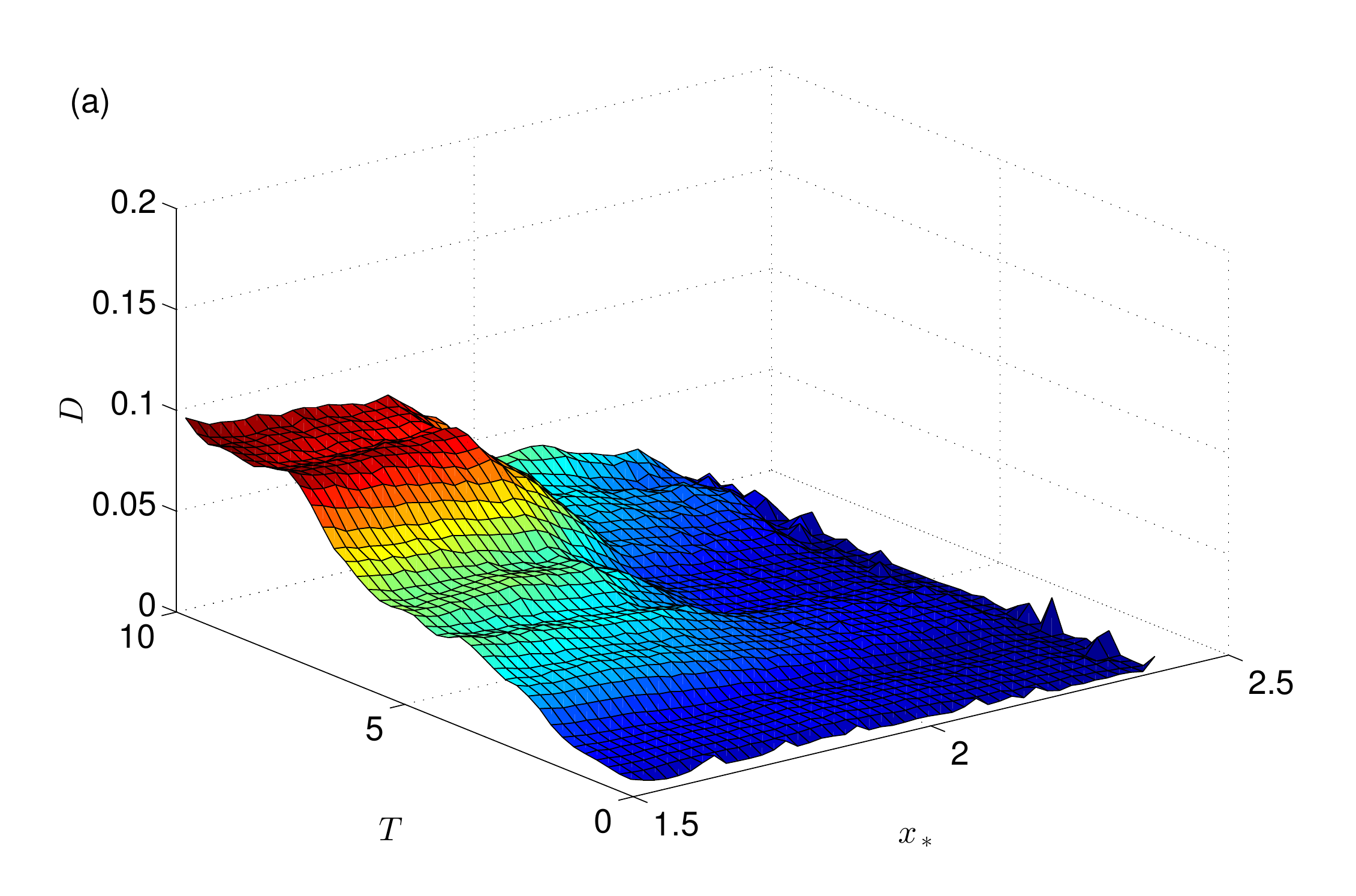} &
            \includegraphics[width=0.5\linewidth]{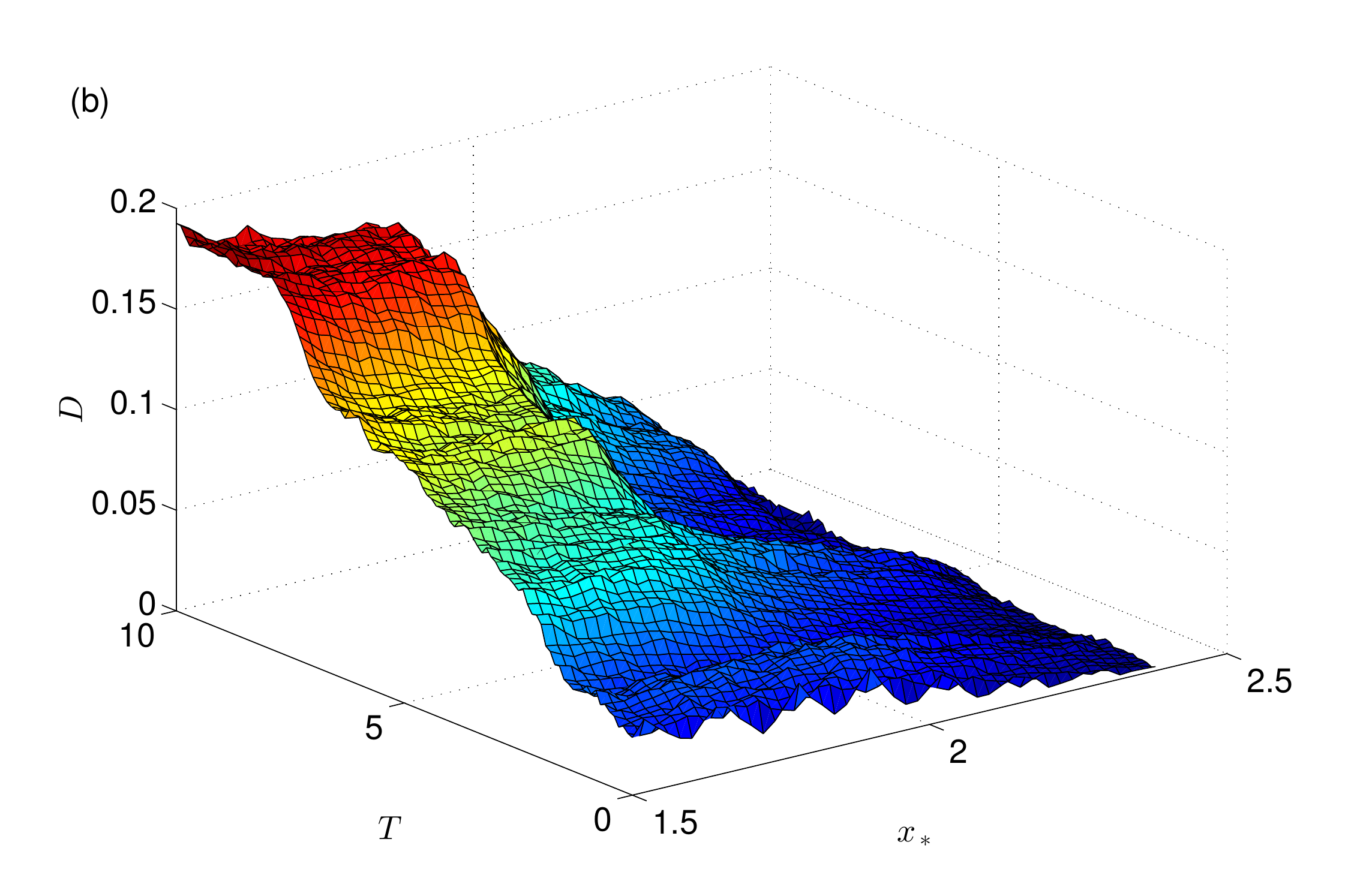} \\
            \includegraphics[width=0.5\linewidth]{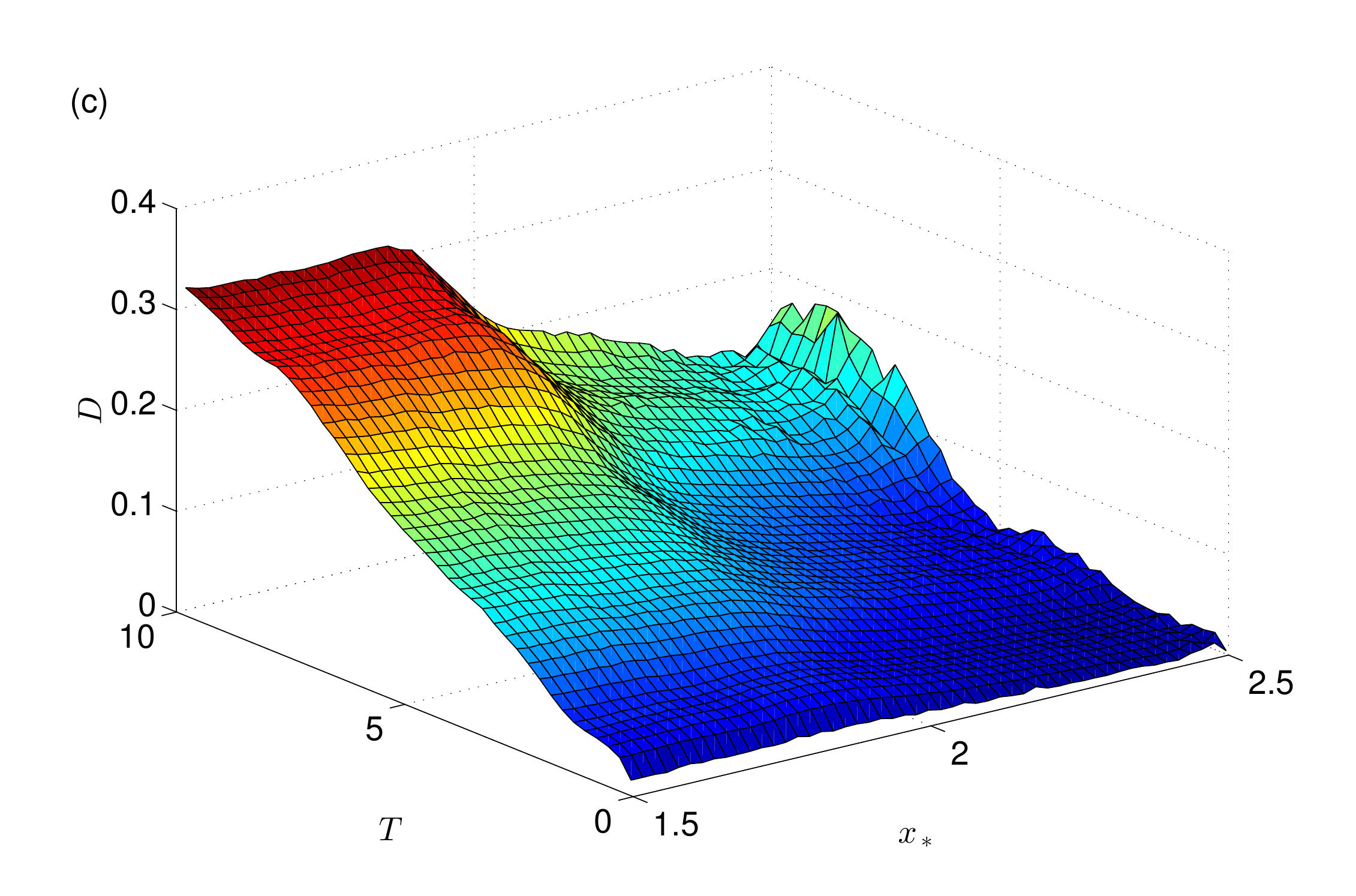} & 
            \includegraphics[width=0.5\linewidth]{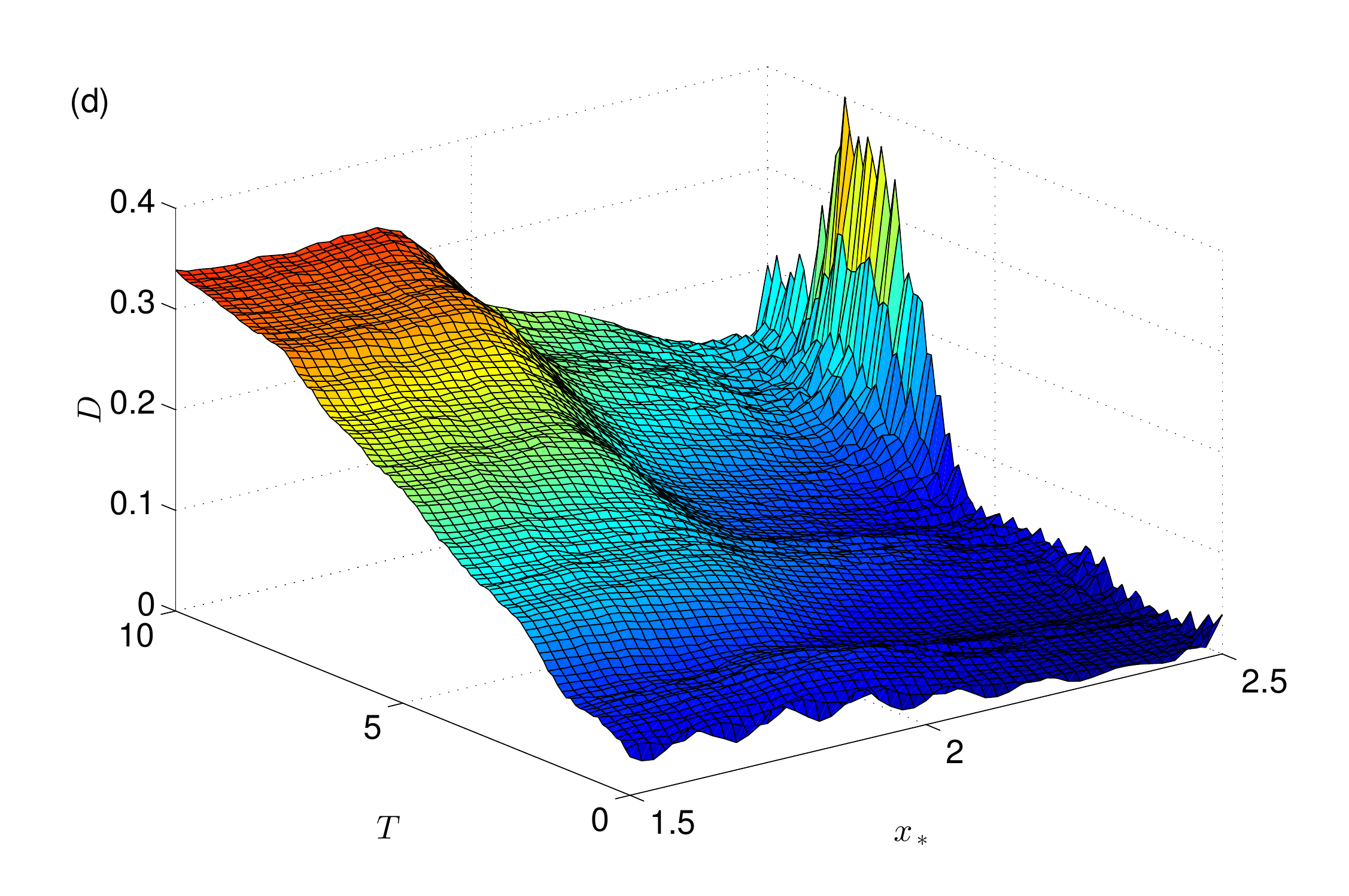}
        \end{tabular}
        \caption{\label{fig:opt_bin} 
        {\hll Predictability in terms of $D$ as a function of the prediction lead time $T$ and threshold level $x_*$. Results are shown for {\hl models M1 and M4 in panels (a)-(b) and (c)-(d), respectively.} The pair of panels in each column correspond to panels (a) and (b) of Fig. \ref{fig:ROC_vs_ftle}} but (a), (c) having the histogram bin size [uniform wrt. one histogram but different for each combination of ($T,x_*$)] optimized, and (b), (d) with $B^{1/3}=25$ bins in all dimensions. Notice the different ranges of $D$ shown.}
    \end{center}
\end{figure*}

{\hl However, Fig.} \ref{fig:ROC_vs_ftle} (b) of the noisy L84 admits values of $D>\sqrt{2}/2$, which should be erroneous. In fact the reason for this error is that the corresponding ROC curves or staircases (not shown) do not extend to the corner (1,1), or more precisely, they feature an excessively large last step -- for the reason stated in the end of {\hll Sec. \ref{sec:numerics}}. This is so because the back-traced data points from the event volume are spread out in relatively large domains of the phase space due to the relatively large prediction lead times and strong instabilities (on average) of the trajectories. As also mentioned already, there is a unique {\em optimal} (uniform) bin size yielding a minimal $D$. We determined numerically this optimum for each sample combination of $(T,x_*)$ separately. This was done using a simple algorithm of maximum finding detailed in Appendix \ref{sec:max_find_alg}, which is suitable for treating nonsmooth functions of one variable. The result of this {\hl optimization} for {\hl M1 and M4} %the undriven and WN-driven L84 
can be seen in Fig. \ref{fig:opt_bin} (a) and (c), respectively. The surprising outcome with the bin size optimization is that stronger events are generally better predictable, reinstating the rule found by Hallerberg and Kantz~\cite{npg-15-321-2008} for the simple stochastic process of AR(1). Only {\hl in case of M4} %the WN-driven case 
do we see an anomaly for very high thresholds, {\hll which could well be the same undersampling effect what was seen in Fig. \ref{fig:d_vs_ths_pot}.} A further observation is that the nonmonotonic $T$-dependence is suppressed/gone almost completely for {\hl M1/M4}. %the undriven/WN-driven L84. 

\begin{figure}  [t!]
    \begin{center}
        \begin{tabular}{c}
            \includegraphics[width=\linewidth]{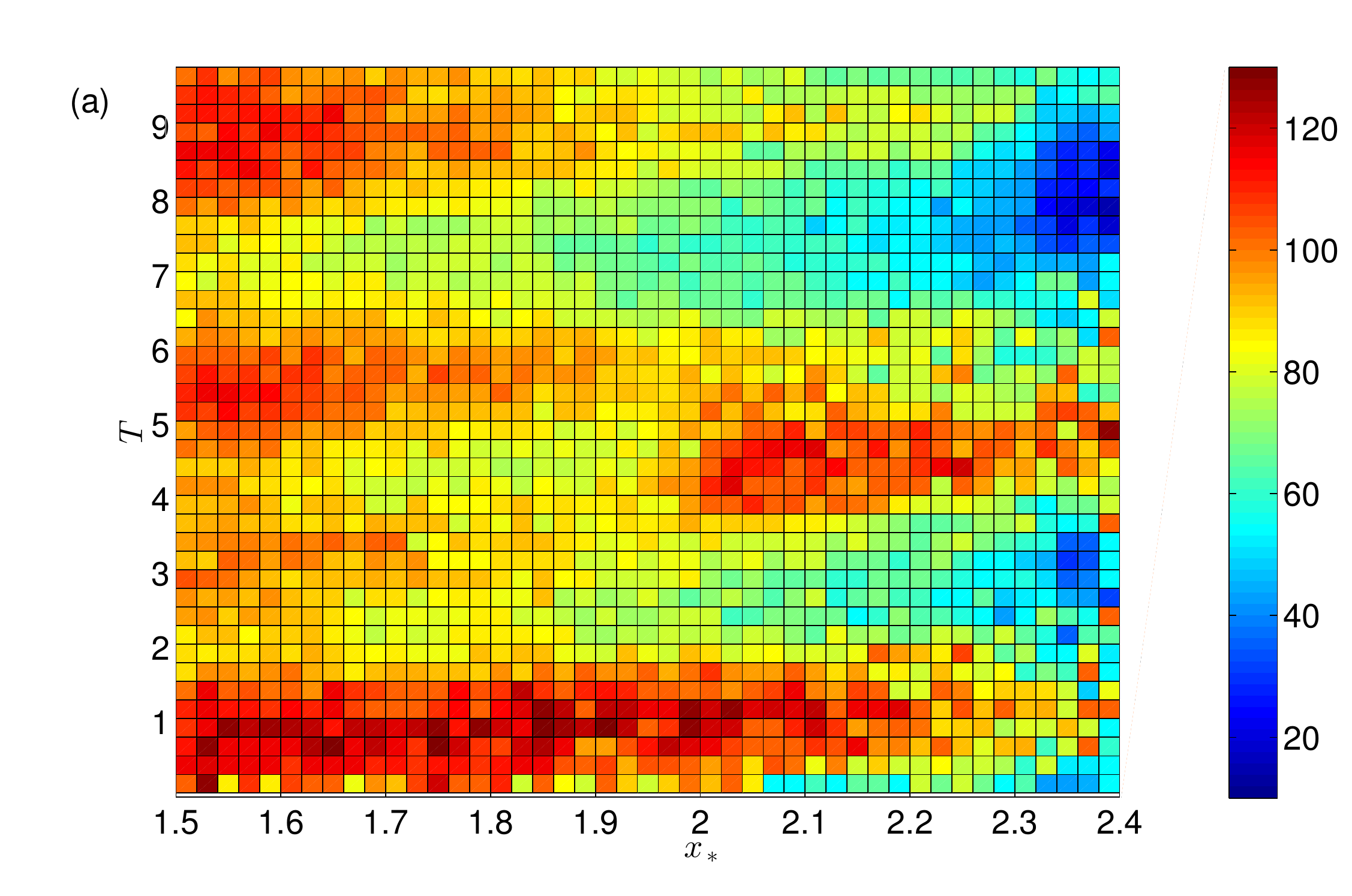} \\
            \includegraphics[width=\linewidth]{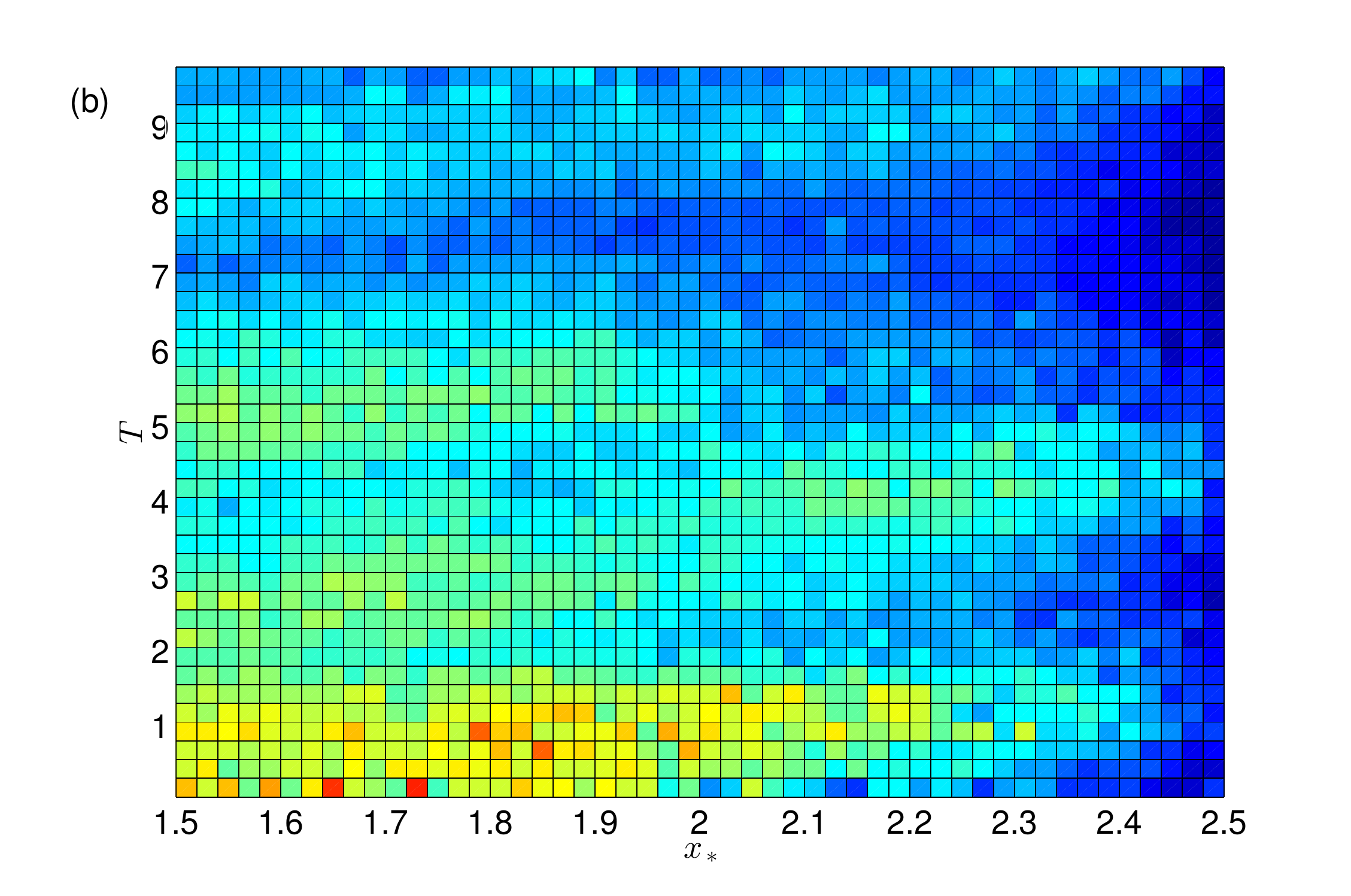}
        \end{tabular}
        \caption{\label{fig:opt_bin2} Optimal number of bins ($B^{1/3}_{opt}$). Panels (a) and (b) {\hll belong} to (a) and (c) of Fig. \ref{fig:opt_bin}. The colorbar applies to both diagrams.}
    \end{center}
\end{figure}

The optimal number of bins are shown in Fig. \ref{fig:opt_bin2}. Comparing these diagrams with the corresponding ones in Fig. \ref{fig:ROC_vs_ftle} one can notice that larger values of {\hll the unoptimized} $D$ correspond to fewer and so larger {\hll optimal} bins. The reason for this is that in these situations the trajectories are more unstable and therefore they scatter in a larger volume, which `asks for' increasing the bin size in order to have a better estimate of the likelihood in those bins.

Finally, we note that MDP %model-driven predictability 
cannot involve such optimization; the bin size is determined by the precision of observation only. {\hll Nevertheless, the unoptimized result in} Fig. \ref{fig:ROC_vs_ftle} does not represent model-driven predictability {\hll either}, because in many bins there is an insufficient number of points for the evaluation of the likelihood. For a given data set size the likelihood can be well-approximated in most bins with a bin size larger than the optimal one for DDP. For the data set size in our analysis we evaluate the model-driven predictability as a function of the prediction lead time $T$ and the threshold level $x_*$ for a `hypothetical' observational precision that derives from approximately the largest optimal bin size in the considered ranges of $T$ and $x_*$, taken to be $B^{1/3}=25$ bins in all three dimensions. The interesting result is that also the model-driven predictability is the better the stronger the extremes. %This statement fails only in case of the WN-driven L84 in a relatively small regime of the strongest extremes when $T>5$, likely because the bin size is not larger than the optimal one. 17.08.15. THIS AGAIN SHOULD BE DUE TO UNDERSAMPLING.

\section{Summary and discussion}\label{sec:summary}

\begin{table*} 
\caption{{\hl Summary of how different factors/choices influence predictability {\hll measured by $D$ (\ref{eq:distance})}. Where appropriate, the mechanism is specified. {\hll Wherever not explicitly specified, the point applies to both POT and TEI events. Whether a point applies to data- or model-driven prediction, or both, should be clear from the context.} A positive/arbitrary/negative effect on predictability is denoted by the symbol $\uparrow/\updownarrow/\downarrow$. We denote the ceiling function by $\lceil\cdot\rceil$.}
}\label{tab:summary}
%\small
\scriptsize
\begin{center}
\begin{tabular}{ L{3cm} | L{6cm} | L{6cm} }
\toprule
{\bf Factor/choice} &
{\bf Effect/mechanism} &
{\bf Support (proof/example)} \\ 
\midrule%\cmidrule{2-3}
0. $\mathcal{L}$-ROC vs $\mathcal{P}$-ROC &
$\uparrow$ $\mathcal{L}$-ROC always better &
\cite{HBK:2008,10.1371/journal.pone.0111506}; Fig. \ref{fig:Pp_L_xx_xphi} (b) vs (c), (e) vs (f), (h) vs (i), (k) vs (l); Fig. \ref{fig:ROC_all}   \\
\midrule%\cmidrule{2-3}
\multirow{4}{*}{
% http://tex.stackexchange.com/questions/187299/position-figure-to-absolute-middle-in-multirow
\raisebox{-35pt}{
% http://tex.stackexchange.com/questions/36056/text-wrap-in-tables-multirow-package-loaded
\parbox{2.8cm}{
1. Makeup of the precursory structure (PS) (assuming other factors fixed)}}} &
1.1 Size of PS $M$ \newline
Extending PS with additional observable, $M'=M+1$: \newline
$\uparrow$ Always better
 &
Embedding theorem~\cite{Takens:1981} \newline
Compare figures in Tables \ref{tab:distance_values_x} and \ref{tab:distance_values_r} (T\ref{tab:distance_values_x}, T\ref{tab:distance_values_r}), in each column separately, row-wise [disregard row 3 (r3) of T\ref{tab:distance_values_x}].
 \\
\cmidrule{2-3}
 &
1.2 Choice of observables &
 \\
\cmidrule{2-2}
 &
1.2.1 A PS of size $M\geq1$ can outperform another PS of size $M'$, $2\lceil D_0\rceil\geq M'\geq1$ no matter $M'>M$ &
Compare r3 to r4 in T\ref{tab:distance_values_x}, wrt. each column separately. \\
\cmidrule{2-3}
 &
1.2.2 Fixing the size $M\leq2\lceil D_0\rceil$, there are an infinity of PSs with varying performance, and there should be an optimal, best-performing, one. Ideal case: precursory space can fully embed attractor. &
Compare Fig. \ref{fig:Pp_L_xx_xphi} (a) to (g) and (b) to (h); and r3 of T\ref{tab:distance_values_x} to other rows; and cases 11, 12 of Fig. \ref{fig:ROC_all} to other cases in it. \\
\midrule%\cmidrule{2-3}
\multirow{5}{*}{
\raisebox{-70pt}{
\parbox{2.8cm}{
2. Intrinsic properties of the system, i.e., the system itself. Therefore, the listed properties change `in tandem' in general.}}} &
Overall effect: \newline
$\updownarrow$ Arbitrary, since at least one of the intrinsic properties have an arbitrary contribution, and they change in tandem.
 &
In general the effects cannot be demonstrated in isolation (see examples for 2.1.1, 2.2), only if one property can be changed independently (see example for 2.1.2). \\
\cmidrule{2-3}
 &
2.1 Attractor &
 \\
\cmidrule{2-2}
 &
2.1.1 Larger dimension $D_0$: \newline
$\downarrow$ The attractor would look more folded
\newline 
Note that 2.1.1 is not a converse of 1.1, because 2.1.2-3, 2.2 changes along with 2.1.1.
 &
Compare r1 to r2 and r3 to r4 in Fig. \ref{fig:Pp_L_xx_xphi}; and figures in T\ref{tab:distance_values_x}, T\ref{tab:distance_values_r} under M1 to M2-3 in each row separately. Exception: M1 to M4 in r1 of T\ref{tab:distance_values_x}, dominated by other intrinsic factors \\
\cmidrule{2-3}
 &
2.1.2 Geometry: \newline
$\updownarrow$ Can enhance or suppress foldedness \newline
2.1.3 Natural measure (probability distribution) \newline
$\updownarrow$ Can enhance or suppress the significance of foldedness
 &
No such exception in T\ref{tab:distance_values_r} as in T\ref{tab:distance_values_x} wrt. 2.1.1, and note that the changes in 2.2 and 2.1.1 between M1 and M4 are the same. Note that it has no significance that also the PSs for M1 and M4 are not the same. \\
\cmidrule{2-3}
 &
2.2 Stronger instability of trajectories:
\newline
$\downarrow$ Faster spread of ensemble of trajectories
\newline
Note: no effect only in case of $N\rightarrow\infty$ and when PS fully embeds attractor in the same time  &
Compare M2 to M3 in T\ref{tab:distance_values_x} or T\ref{tab:distance_values_r}. MLE of L63 increases for decreasing $\tau'$, but the impact on L84 wrt. either 2.1.2-3 or 2.2 clearly counters and outweighs that (except for r1 in T\ref{tab:distance_values_x}).  \\
\midrule%\cmidrule{2-3}
3. Data set size $N$
(other factors being arbitrarily fixed)
 &
Smaller $N$:
\newline
$\downarrow$ Larger statistical errors in estimating $\mathcal{L}$
 &
Intuitive statement not checked empirically \\
\midrule%\cmidrule{2-3}
4. Precision of measurement $\delta x$ &
Larger $\delta x$: \newline
$\downarrow$ Smoothing or coarser-graining in estimating $\mathcal{L}$
 &
Intuitive statement not checked empirically \\
\midrule%\cmidrule{2-3}
5. Bin size $\Delta x$
($N$ being finite and fixed)
 &
$\updownarrow$ An optimum $\Delta x_{opt}$ exists, which might always be unique. &
Our experience in all cases examined in numerics \\
\midrule%\cmidrule{2-3}
6. Prediction lead time $T$ {\hll (only for TEI events)} &
Increasing $T$:
\newline
$\updownarrow$ Possibly nonmonotonic $D(T)$
 &
Fig. \ref{fig:ROC_vs_ftle} \\
\cmidrule{2-3}
 &
$\downarrow$ Increasing $D(T)$ for $\Delta x_{opt}$ &
Fig. \ref{fig:opt_bin} (a,c) \\
\midrule%\cmidrule{2-3}
7. Event magnitude $x_*$
 & Increasing $x_*$: \newline 
 {\hll POT events: \newline  
 $\uparrow$ $\updownarrow$ $D(x_*)$ decreasing for highest thresholds on coarse scales, but could be nonmonotonic on `fine' scales even for $\Delta x_{opt}$}  & 
 Fig. \ref{fig:d_vs_ths_pot}
 \\
\cmidrule{2-3}
 &
{\hll TEI events:}
\newline
$\updownarrow$ Possibly nonmonotonic $D(x_*)$ &
Fig. \ref{fig:ROC_vs_ftle}
 \\
\cmidrule{2-3}
 &
$\uparrow$ Decreasing $D(x_*)$ for $\Delta x_{opt}$ &
Fig. \ref{fig:opt_bin} (a,c) \\
\bottomrule%\cmidrule{2-3}
\end{tabular}
\end{center}
\end{table*}

\begin{figure} [t!]
    \begin{center}
	  \includegraphics[width=\linewidth]{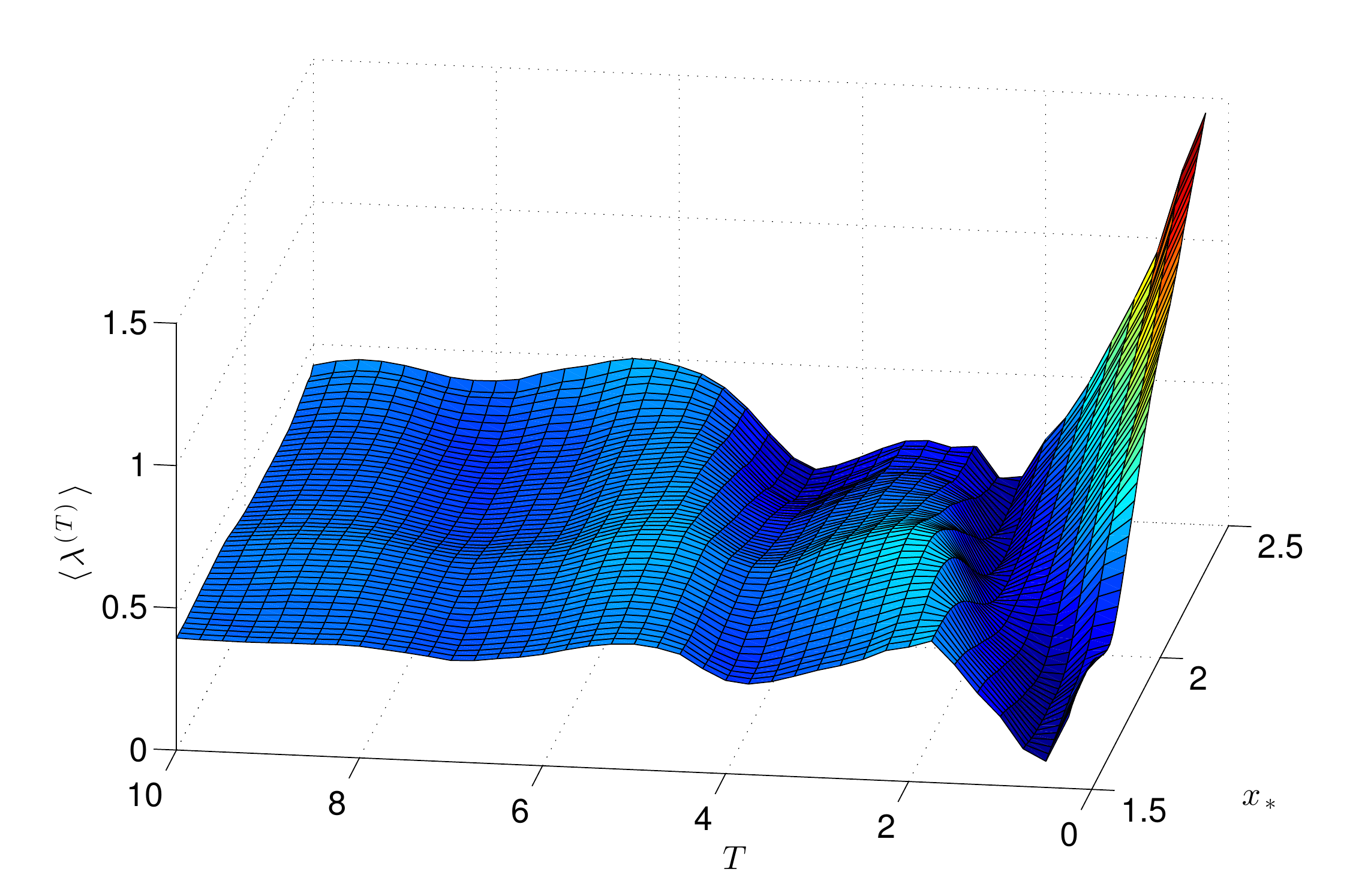} 
        \caption{\label{fig:ftle_vs_xth_T} Predictability in terms of the average finite-time maximal Lyapunov exponent $\langle\lambda^{(T)}\rangle$ as a function of the prediction lead time $T$ and threshold level $x_*$, corresponding to Fig. \ref{fig:ROC_vs_ftle} (a). {\hl The exact correspondence is established by averaging the FTMLEs over the event volume of TEI events. The latter is defined in Sec. \ref{sec:lead_time}, and the corresponding average is formulated by Eq. (\ref{eq:mean_ftmle_DT}).}}
    \end{center}
\end{figure}

We examined the predictability of threshold exceedance `extreme' events in a simple but chaotic continuous-time dynamical system or `flow'. Given the nature of the problem, namely, that extremes are rare, we chose an arguably~\cite{Kantz:2010} appropriate {\hl measure of prediction skill} for assessing predictability: the ROC statistics, more specifically, a distance measure $D$ from the ideal situation of having all events successfully predicted without any false alarms. {\hl According to our top objective (i) {\hll and (iii)} set out in the Introduction,} we examined the dependence of predictability on various factors. {\hl Our conclusions are collected in a systematic form in Table \ref{tab:summary} {\hll(T\ref{tab:summary})}; in the column on the right we refer to our results presented in this paper, and previous results reported by others, to support the statements in the middle column.

Among the factors we did not list the choice of observable for $x$ {\hll whose extremal values are concerned}. The reason for this is that we cannot make a statement of general interest with respect to this choice. The predictability of extremes concerning different observables $x$ and $x'$ can be compared only if we name conditions that have to be satisfied in both cases. Such a condition on the threshold levels $x_*$ and $x'_*$ can hardly be given objectively; one possibility is that the threshold level should belong to the same quantile of the respective process distributions $p(x)$ and $p(x')$. However, this is still a subjective condition. And since the predictability strongly depends on the threshold, a comparison is hardly possible\footnote{This argument can be applied also to the intrinsic properties, i.e., the choice of the system in general (point 2 in T\ref{tab:summary}). However, it can be of general interest to compare the predictability with respect to the same physical observable while only slightly changing the system, e.g. by changing a parameter, or by considering different types of perturbations as illustrated by our model choices M1-4.}.   

Instead of merely the distance $D$, we can compare the monotonicity or trend (increasing or decreasing) of $D(x_*)$ and $D(x'_*)$. {\hll This is the difference between objectives (i.a) and (i.b).} The interesting finding in this regard is that $D(x_*)$ is monotonically decreasing {\hll concerning TEI events} -- although only if the bin size of histograms is optimized (see point/conclusion 7 (c7) in T\ref{tab:summary}). {\hll Concerning POT events the situation is somewhat more intricate, but on coarse scales of the threshold level we observe the same behavior. It appears to be robust, being the same qualitatively for two different precursory structures and for all of M1-4.} Since we found this effect in an arbitrarily chosen dynamical system, and with respect to an arbitrarily chosen observable of it, we suggest that it might be a rather typical behavior. This would be a nontrivial generalization of the same statement made by Hallerberg and Kantz~\cite{npg-15-321-2008} concerning autoregressive processes. A theoretical argument why this should or {\hll rather} should not be always true is yet to be provided, however.

We point out that the above observation was made in the special case, {\hll among other cases,} when the precursory space is identical to the phase space. {\hll There was one case concerning POT events (to do with the gray lines in Fig. \ref{fig:d_vs_ths_pot}), and another one concerning TEI events (see e.g. Fig. \ref{fig:opt_bin}).} Therefore, the effect of folding referred to under points 2.1.1-3 of T\ref{tab:summary} does not take place. However, we have already seen evidence that even in this case this is not the stability of trajectories (point 2.2 of {\hll T\ref{tab:summary}}) alone, or not that measured by the average finite-time maximal Lyapunov exponent $\langle\lambda^{(T)}\rangle$ (defined in Appendix \ref{sec:pred_model_driv}), that determines $D$: As seen in Figs. \ref{fig:ROC_vs_ftle} (a) and \ref{fig:opt_bin} (b), the bin size alone can change the monotonicity or trend of $D(x_*)$, while $\langle\lambda^{(T)}(x_*)\rangle$ is obviously unchanged. The latter is shown in Fig. \ref{fig:ftle_vs_xth_T}. This mismatch is not a finite data set size numerical effect, as $\lim_{\Delta x\rightarrow0}\lim_{N\rightarrow\infty}D=1$ with no $(T,x_*)$-dependence that could match that of $\langle\lambda^{(T)}\rangle$. We might say rather that (assuming $N\rightarrow\infty$ {\hll for DDP, or that we concern on-demand MDP}) $\Delta x$ {\hll(or $\delta x$)} controls the `filtering' of intrinsic properties in determining $D$. We note that, {\hll provided that $\Delta x>0$ (or $\delta x>0$)}, a similar {\hll filtering} role can be played also by $x_*$. {\hll However, some other role seems to be played by the process PDF $p_x(x_*)$ too (see Fig. \ref{fig:d_vs_ths_pot} (b)).} With this, we believe to have reached our objective (iv).}

Advanced algorithms to bin data will be the objective of our future research. In this regard it is envisaged that a theory linking {\hl formally some suitable measure of the instability of trajectories} %the Lyapunov exponent(s) 
etc. with the prediction skill {\hl in terms of $D$} could indicate the optimal grid in a straightforward manner, rather than having to find this grid by conducting a costly general iterative optimization procedure. Furthermore, it is often the case that the data set is very limited for pure data-driven prediction, while some model, even if inaccurate at the current stage of its development, is known. In this case it would be most beneficial if a combined data- and (archival or on-demand) model-driven prediction technique could exploit fully the assets of data and model at hand. We will concentrate efforts to develop such techniques. 

{\hl Finally we remark that point (2) of the Introduction does seem to contradict our conclusion (c7). This is so given that we argued in {\hll Sec. \ref{sec:formal_setting}} -- according to our objective (ii) -- that (c7) should apply also to model-driven predictions. However, it might be crucial to assume no model errors in order to reach (c7). This assumption certainly never holds in practice, and it might indeed lead to point (2) to hold. Model errors can be easily modeled, say, by taking M3 as the truth, and M4 as the model. Another reason for the apparent contradiction between (2) and (c7) may also be that they assume different measures of prediction skill. In order to possibly reach completely our objective (ii), we should- and we wish to examine these issues in a separate work.}

\section*{Acknowledgments}

Much of this work was carried out at the Max Planck Institute for the Physics of Complex Systems, Dresden, Germany; the scholarship awarded by the Max Planck Society is gratefully acknowledged. The work was supported also by the NAMASTE project owned by Valerio Lucarini (under the ERC grant No. 257106). The author is indebted for useful discussions with Holger Kantz, Tam\'as T\'el, and Christian Franzke. Useful comments on the manuscript by Frank Lunkeit are gratefully acknowledged. The author is thankful to two anonymous reviewers for many helpful suggestions to improve this work.



\begin{appendices}

\section{Algorithm for finding the approximate global maximum of a nonsmooth function of one variable}\label{sec:max_find_alg}

The ROC{\hl-based measure of prediction skill} $D$ is a discontinuous function of the linear bin size $\Delta x$ due to the finite data set size $N$. Conversely, if there was infinite data available, it would be a continuous function. We will assume here that with finite $N$, $D$ features a single minimum. If a function is discontinuous or nonsmooth, the Newton-Raphson algorithm that relies on the derivative cannot be applied to find a global minimum. 

Instead of the bin size $\Delta x$, we will specify the number $B$ of bins {\hll (or $B^{1/n}$ along a single dimension in the $n$-dimensional precursory space)} in the domain where data points are found. Our experience is that, given the data available, as specified in Sec. \ref{sec:model}, and the ranges of $T$ and $x_*$ desired to be explored, the optimal number $B_{opt}^{1/3}$ of bins is between, say, 6 and 200. One could evaluate $D$ for all intermediate integers to find the one that gives the smallest $D$. However, one can do better than that. The following algorithm is applicable to smooth functions $f(x)$ possessing a single maximum, but also to discontinuous/nonsmooth or discrete approximations of such functions, provided that the root-mean-square error of approximation is relatively small (loosely speaking: smaller than the `elevation of the maximum'). To start with, we define five equally spaced values of the independent variable $x\in\mathbb{R}$ determined by the choice for the smallest and largest values: $x_{i,j}$, $i=1,\dots,5$, $x_{2/3/4,j}=(x_{1/1/3,j}+x_{3/5/5,j})/2$, $j$ being the iteration variable. (A rounding can be applied if integer values of $x_{i,j}$ are accepted only.) Initially we set $x_{1,0}=6$ and $x_{5,0}=200$. Then in each iteration, $j=1,2,\dots$, we check the following cases: 

\begin{description}
 \item[Case 1] $\max_i[f(x_{i,j})]=f(x_{3,j}) \rightarrow$\\$ x_{1,j+1}=x_{2,j},\ x_{5,j+1}=x_{4,j}$  
 \item[Case 2] $f(x_{3,j})>f(x_{2,j})\ \&\ f(x_{4,j})>f(x_{3,j}) \rightarrow$\\$ x_{1,j+1}=x_{3,j},\ x_{3,j+1}=x_{4,j}$
 \item[Case 3] $f(x_{3,j})<f(x_{2,j})\ \&\ f(x_{4,j})<f(x_{3,j}) \rightarrow$\\$ x_{5,j+1}=x_{3,j},\ x_{3,j+1}=x_{2,j}$
 \item[Case 4] otherwise $\rightarrow$\\
 $\max[f(x)]\approx x_{3,j}$
\end{description}

Case 4 is never encountered in case of a smooth function featuring a single maximum, and the iteration would go on indefinitely without a stopping condition. Considering discontinuous/nonsmooth or discrete approximants of such functions the iteration is terminated in finite time ($j$).

\section{Finite-time Lyapunov exponent}\label{sec:pred_model_driv}

A well-known measure of predictability is the positive maximal Lyapunov exponent (MLE), which approximates the average rate of the exponential separation of very close trajectories on a chaotic attractor~\cite{Tel_n_Gruiz:2006}. Sterk et al.~\cite{npg-19-529-2012} evaluated the finite-time version of this measure to assess the predictability, with some lead time $T$, of extreme events. We consider here `apriori' known nonautonomous dynamical systems $\dot{y}=f(y,t)$ in a $d$-dimensional phase space with generic initial condition $y_0=y(y_0,t=t_0,t_0) \in \mathbb{R}^d$, where $y(\cdot,\cdot,\cdot)$ denotes the two-time evolution operator. 

The spectrum of finite-time $T$ Lyapunov exponents (FTLE) $\lambda_i^{(T)}$ can be defined in a {\em pullback} sense~\cite{PhysRevE.87.042902} as follows:

\begin{equation}\label{eq:ftle}
  \lambda_i^{(T)} = \lim_{t_0\rightarrow-\infty}\frac{\ln(s_i^{1/2}(t,t_0))-\ln(s_i^{1/2}(t-T,t_0))}{T},
\end{equation} 
$i=1,\dots,d$, where $s_i(t,t_0):\ \det(Y(t,t_0)\cdot Y^T(t,t_0) - sI)=0$ are the singular values of the deformation gradient, $Y=\partial y/\partial y_0$, governed by the variational equation:

\begin{equation}\label{eq:var_eq}
  \dot{Y} = \left.\frac{\partial f}{\partial y}\right|_{y(y_0,t,t_0)}\cdot Y.
\end{equation}
The initial condition is not arbitrary but implied as $Y(y_0,t_0,t_0)=I$. Note that the LEs are recovered as: $\lambda_i=\lim_{T\rightarrow\infty}\lambda_i^{(T)}$. We will omit the index $i$ to denote the MLE simply by $\lambda$ or $\lambda^{(T)}$. Clearly, by $\lambda^{(T)}(t)$ the predictability from the present time $t-T$ of a trajectory at the future time $t$ is defined. {Note that by the inversion $t=t(y,t_0,y_0)$ we have $\lambda^{(T)}(y,t_,y_0)$.}

A summary statistics for this measure of predictability can be defined, generalizing the proposal of Sterk et al.~\cite{npg-19-529-2012}, by an ensemble average over parts of the pullback or {\em snapshot attractor}~\cite{PhysRevE.87.022822} that realize extreme events in terms of some physical observable $x(y)$: 

\begin{equation}\label{eq:mean_ftmle}
\begin{split}
  \langle\lambda^{(T)}(t)\rangle = &\int\mu(y,t) dV_y\lambda^{(T)}(y,t)\times \\
  &\mathcal{H}(x(y)-x_*),
\end{split}
\end{equation}
{\hl where $\mu(y,t)$ is the natural measure supported by the snapshot attractor~\cite{DBT:2015}}. {\hll Alternatively, the average can be taken} over parts of the snapshot attractor collecting the ensemble of trajectories that would cross the threshold in a leading window of time of width $\Delta T$ at time $t$ (TEI events):

\begin{equation}\label{eq:mean_ftmle_DT}
\begin{split}
  %\langle\lambda^{(T)}(t)\rangle = \int_{x(y(\eta,\tau=t,t))=x_*}^{x(y(\eta,\tau=t+\Delta T,t))=x_*}\frac{d\eta}{d\tau}d\tau\lambda^{(T)}. 
  \langle\lambda^{(T)}(t)\rangle = &\int\mu(y,t) dV_y\lambda^{(T)}(y,t)\times \\ 
  &\mathcal{H}(x(y)-x_*)\times \\
  &\mathcal{H}(x_*-x(y_0(y,t,t-\Delta T))),
\end{split}
\end{equation}
{\hl where $y_0(y,t,t_0)$ is obtained by the inversion of the two-time evolution operator}. The average FTLE or FTMLE $\langle\lambda^{(T)}\rangle$ is dissimilar to $D$ in that it is not calculated from predicted data, but rather it expresses an intrinsic property of the system that determines predictability. Nevertheless, we will compare figures obtained for $\langle\lambda^{(T)}\rangle$ by (\ref{eq:mean_ftmle_DT}) and $D$, at least in case of the autonomous dynamics when the attractor is time-invariant. 

\end{appendices}

% If you have bibdatabase file and want bibtex to generate the
% bibitems, please use
%
%  \bibliographystyle{elsarticle-num} 
%  \bibliography{../predict_extremes}


\end{document}